\documentclass[prd,aps,nofootinbib,eqsecnum,notitlepage,nobibnotes,superscriptaddress,preprintnumbers,floatfix,twocolumn]{revtex4-1}

\usepackage{subfig}
\usepackage[a4paper, left=20mm,right=20mm,top=20mm,bottom=20mm]{geometry}
\usepackage[titletoc,toc,title]{appendix}
\usepackage{mathtools}
\usepackage{tensor}
\usepackage{amsmath, amsfonts,bm,amssymb}
\usepackage{derivative}
\usepackage{bm}
\usepackage{suffix}
\usepackage{soul}
\usepackage{amsmath, amsfonts,bm}
\usepackage{color}
\usepackage{graphicx}
\usepackage[usenames,dvipsnames]{xcolor}
\usepackage{subfig}
\usepackage{ulem}
\definecolor{myred}{rgb}{0.9, 0.17, 0.31}

\usepackage[a4paper, left=20mm,right=20mm,top=20mm,bottom=20mm]{geometry}
\usepackage[colorlinks=true,hyperfootnotes=true, linkcolor=myred, citecolor=cyan, urlcolor=magenta]{hyperref}

\newcommand\mathcomma{\,,}
\newcommand\mathperiod{\,.}

\DeclareMathAlphabet{\mathup}{OT1}{\familydefault}{m}{n}

\def\dd{\mathrm{d}}

\renewcommand\vec[1]{\bm{#1}}

\newcommand{\be}{\begin{equation}} 
\newcommand{\ee}{\end{equation}}

\begin{document}

\title{Coupling quintessence kinetics to electromagnetism}

\author{Bruno J. Barros}
\email{cstbru002@myuct.ac.za}
\affiliation{Cosmology and Gravity Group, Department of Mathematics and Applied Mathematics, University of Cape Town, Rondebosch 7700, Cape Town, South Africa}
\author{Vitor da Fonseca}
\email{vitor.dafonseca@alunos.fc.ul.pt}
\affiliation{Instituto de Astrof\'isica e Ci\^encias do Espa\c{c}o,\\ 
Faculdade de Ci\^encias da Universidade de Lisboa,  \\ Campo Grande, PT1749-016 
Lisboa, Portugal}

\date{\today}

\begin{abstract}

We propose a general model where quintessence couples to electromagnetism via its kinetic term. This novelty generalizes the linear dependence of the gauge kinetic function on $\phi$, commonly adopted in the literature. The interaction naturally induces a time variation of the fine-structure constant that can be formulated within a disformally coupled framework, akin to a Gordon metric. Through a suitable parametrization of the scalar field and the coupling function, we test the model against observations sensitive to the variation of $\alpha$. We undertake a Bayesian analysis to infer the free parameters with data from Earth based, astrophysical and early Universe experiments. We find that the evolution of $\alpha$ is specific to each cosmological era and slows down at late times when dark energy accelerates the Universe. While the most stringent bound on the interaction is obtained from atomic clocks measurements, the quasars provide a constraint consistent with weak equivalence principle tests. This promising model is to be further tested with upcoming and more precise astrophysical measurements, such as those of the ESPRESSO spectrograph.
\end{abstract}

\maketitle

\section{Introduction}\label{sec:intro}

In 1916, German physicist Arnold Sommerfeld introduces the enigmatic {\it fine-structure constant} \cite{Sommerfeld1916}, $\alpha$, to improve Bohr atom's model and explain the splitting of spectral lines in hydrogen into closely spaced (fine) lines. By taking into account relativistic corrections and elliptical orbits of electrons, new energy levels emerge corresponding to the fine-structure of the spectral lines. The origin of the specific value of $\alpha$ remains a mystery and has captured the attention of many renowned physicists, such as Feynmann and Pauli \cite{sherbon:hal-02281150}. More recently, a theoretical effort has been devoted to fundamentally explain the asymptotic low-energy value of $\alpha$ \cite{Singh_2021}. The constants of nature play a crucial role in the description of natural phenomena, such as setting the critical scales for the fundamental forces. One would expect their values to vary in the early Universe, or more specifically, at high energies.

The constancy of the fundamental constants has been a subject of discussion for almost a century (see a detailed review in \cite{Uzan:2010pm}). It is usually traced back to Dirac, 1937, in his ``Large Numbers hypothesis" \cite{Dirac:1937ti,Dirac:1938mt}, where he suggests that such numbers cannot be just arbitrary entities entering the equations, but must vary in time whilst related to some cosmological state. Six months after Dirac's numerological work, Jordan \cite{Jordan1937} proposed a varying $G$ model framed within a field theory approach. This work served as a motivation for the seminal paper of Brans and Dicke \cite{Brans:1961sx} in the sixties, which later led to the modern so called scalar-tensor theories of gravity \cite{Horndeski:1974wa,Clifton:2011jh}. Twenty years pass, until Bekenstein formulated an electromagnetic field theory which allows for variations of the fine-structure constant by varying the electric charge $e$ \cite{Bekenstein:1982eu}. By promoting $e=e_0\epsilon(x^{\mu})$ one notices that $\epsilon$ - function of the spacetime coordinates - couples to the electromagnetic gauge field as $\epsilon A_{\mu}$ rendering the electromagnetic action invariant under the symmetry ${\epsilon A_{\mu}\mapsto \epsilon A_{\mu} + \chi_{,\mu}}$ \cite{Sandvik:2001rv,Bekenstein:2002wz}. The most interesting features emerge when one identifies $\epsilon (x^{\mu})$ as a physical field, endowing it with dynamics through a suitable choice of the Lagrangian density $\mathcal{L}_{\epsilon}$. One year after Bekenstein's work, it was shown \cite{Marciano:1983wy} that if the mean Kaluza-Klein radius of the extra-dimensions is allowed to vary in time, it can induce variations on the fundamental constants within our 4D world. This might provide a method to probe extra-dimensions. Naturally, one could evoke the anthropic principle to state that if any variation in $\alpha$ exists in nature it should not be large, otherwise stable matter and intelligent life as we know it would be impossible to form \cite{BARROW2001}. Nonetheless, observations still leave room to accommodate, albeit small, putative spacetime variation of the fine-structure constant.

A number of studies motivated by Bekenstein's work consider as a starting point a coupling mediated by a field $\phi$ that drives both the late time cosmic acceleration and the time variation of the fine-structure constant \cite{martins2017}, {\it i.e. $\epsilon\rightarrow\phi$}. This kind of models belong to the so-called Class I, in opposition to Class II in which the scalar degree of freedom does not supply the dark energy component. Up to this day, no interactions between quintessence and the Standard Model of particle physics have been observed. This fact must imply the existence of processes suppressing such couplings at tree level \cite{Carroll:1998zi}. However, it still remains possible to accommodate couplings between a slow-rolling scalar field and electromagnetism. Here, as a novelty, we assume that the interaction term $h=h(\phi,X)$, referred to as gauge kinetic function, also depends on the derivative of the field, more precisely, on the kinetic term ${X=-1/2g^{\mu\nu}\partial_{\mu}\phi\,\partial_{\nu}\phi}$, beyond the value of the field {\it per se}. The kinetics of the quintessence field could reasonably contribute to the coupling with the electromagnetic sector since what characterizes the cosmic acceleration is the slow-roll condition in some potential $V(\phi)$ that applies to the kinetic term, ${X\ll V}$. 

We will show how this coupling formalism can be obtained by assuming that photons experience a different metric, $\tilde{g}_{\mu\nu}$, related to $g_{\mu\nu}$ through a disformal transformation \cite{vandeBruck:2015rma}. $\tilde{g}_{\mu\nu}$ corresponds to a Gordon metric \cite{Gordon1923} that allows us to model the gauge kinetic function as the effective refractive index of the scalar field viewed as an optical medium. How dark energy might give rise to $\alpha$ variations through disformal transformations was already noticed in Ref.~\cite{Parkinson:2003kf}. In a model independent approach and resorting to constraints from the weak equivalence principle (WEP), the authors show that spacial perturbations in $\alpha$ need to be small and that these can be related with perturbations of the dark energy field.

This paper is organized as follows. In Sec.~\ref{sec:model}, we describe the original model for varying $\alpha$ that we propose. It includes our choice of parametrizations, for the scalar field and the gauge kinetic function. The cosmological behavior of $\alpha$ is derived in each era dominated by the different energy sources. We test the model in Sec.~\ref{sec:obs} and present the constraints obtained on the free parameters from the various datasets used. The results are wrap-up in Sec.~\ref{sec:conclusions} where we discuss notable features of the model.

\section{An original model for varying $\alpha$}\label{sec:model}

\subsection{Coupling $\phi$ to electromagnetism}\label{model_intro}

We consider the total action of standard Einstein gravity, as well as a scalar field coupled to the strength tensor $\bf{F}=\bf{dA}$, with components ${F_{\mu\nu}=2\partial_{[\mu}A_{\nu]}}$, where $A_{\mu}$ is the electromagnetic four-potential of classical electromagnetism,
\begin{eqnarray}
\mathcal{S} = \int \dd^4x\sqrt{-g}\bigg[&\,&\frac{1}{2\kappa^2}R+\mathcal{L}_{\phi}-\frac{1}{4\mu_0}h(\phi,X)F^{\mu\nu}F_{\mu\nu} \nonumber \\
&\,& -A^{\mu}j_{\mu}  \bigg]+\mathcal{S}_m\mathcomma\label{action}
\end{eqnarray}
where $\dd^4x\sqrt{-g}$ is the volume form, $g$ being the determinant of the metric tensor $g_{\mu\nu}$, ${\kappa^2=8\pi G/c^4}$, $R$ is the curvature scalar, $\mathcal{L}_{\phi}$ is the scalar field Lagrangian density, $j^{\mu}$ are the components of the free four-current, $\bf{J}$, sourced by standard model charged particles, $\mu_0$ is the permeability of vacuum, and $\mathcal{S}_m$ collectively stands for the action of matter fields.

It is usually assumed that the interaction between the scalar field and the electromagnetic sector results from a coupling to the kinetic term, $F^{\mu\nu}F_{\mu\nu}$, of the gauge field, where the coupling function $h$ is assumed to be linear on $\phi$ in a first-order approximation \cite{martins2017}. The originality of our approach is the gauge kinetic function that explicitly depends on both the value of the scalar field and its kinetic energy, $h(\phi ,X)$. The evolution of the fine-structure constant reads \cite{Damour:1994zq,Damour:2002nv,Copeland:2003cv,martins2017}
\be\label{alpha}
\alpha\propto h^{-1}\mathperiod
\ee
Therefore, the value of the fine-structure throughout the cosmic history depends on the dynamics of the scalar source. In this paper, the latter is identified as a quintessence field permeating the Universe, which drives its current accelerated expansion, by choosing the appropriate Lagrangian density $\mathcal{L}_{\phi}$, as in section~\ref{cosmology}.

Varying the action with respect to the gauge field, $A_\mu$, gives the following electromagnetic field's equations of motion,
\be\label{maxwell}
\nabla_{\mu}\left( hF^{\mu\nu} \right) = \mu_0 j^{\nu}\mathcomma
\ee
or, expanding the covariant derivative,
\be\label{maxwell2}
\nabla_{\mu}F^{\mu\nu}= \frac{\mu_0}{h}j^{\nu} +\left(\frac{h_X}{h}\partial_{\alpha}\phi\nabla_{\mu}\partial^{\alpha}\phi-\frac{h_{\phi}}{h}\nabla_{\mu}\phi \right)F^{\mu\nu}\mathcomma
\ee
where ${h_X=\partial h/\partial X}$ and ${h_{\phi}=\partial h/\partial \phi}$. The emergence of the first term inside brackets in Eq.~\eqref{maxwell2} expresses the influence of the quintessence kinetic term on electromagnetism. In the absence of coupling, by setting ${h=1}$, the last term vanishes and we recover the classical Maxwell's electrodynamics ${{\bf{\nabla}}\cdot{\bf{F}} = \mu_0{\bf{J}}}$ (Bekenstein's first assumption P1 \cite{Bekenstein:1982eu}).

Varying $\alpha$ models can also be framed within Variable-Speed-of-Light (VSL) cosmological models \cite{Barrow:1998df}. Ref.~\cite{Bassett:2000wj} discussed how disformal transformations can be useful in describing the geometrical framework of VSL models, as an alternative to classically promoting ${c\mapsto c(t)}$.

\subsection{$h(\phi,X)$ induced by disformal transformations}\label{disf_subsubsection}

The action in Eq.~\eqref{action} cannot naturally emerge from a purely conformal transformation in the Einstein frame, ${\tilde{g}_{\mu\nu}=C(\phi ,X)g_{\mu\nu}}$, since the trace of the electromagnetic energy-momentum tensor vanishes (radiation is conformally invariant). However, a disformal transformation of the form,
\be\label{disformal_transformation} 
\tilde{g}_{\mu\nu} = C(\phi,X)g_{\mu\nu}+D(\phi,X)\partial_{\mu}\phi\partial_{\nu}\phi\mathcomma
\ee
can achieve such an interaction. Let us assume a theory where radiation experiences a different metric, $\tilde{g}_{\mu\nu}$, disformally related to the gravitational one, $g_{\mu\nu}$, through Eq.~\eqref{disformal_transformation}. While the other components follow the gravitational metric $g_{\mu\nu}$, only radiation is set to follow geodesics of $\tilde{g}_{\mu\nu}$. Accordingly, the action \eqref{action} is changed through the direct mapping,
\be\label{disformal_lagrangian}
\sqrt{-g}\,h(\phi,X)F^{\mu\nu}F_{\mu\nu} \mapsto \sqrt{-\tilde{g}}\Big(\tilde{g}^{\mu\alpha}\tilde{g}^{\nu\beta}F_{\mu\nu}F_{\alpha\beta}\Big)\mathcomma
\ee
where ${\tilde{g}_{\mu\nu}}$ is a function of both $\phi$ and $X$. Varying this action with respect to $A_\mu$, and comparing the result with the time component of Eq.~\eqref{maxwell2} in Minkowski space, we identify \cite{vandeBruck:2015rma}
\be\label{h_disformal}
\sqrt{1-\frac{2DX}{C}}=\frac{1}{h}\mathperiod
\ee
By solving the latter equation in order of $D$, we find the specific disformal transformation that gives rise to the same coupling function, $h$, in our original action \eqref{action},
\be\label{disformal_final}
\tilde{g}_{\mu\nu} = Cg_{\mu\nu}+\frac{C}{2X}\left( 1-\frac{1}{h^2} \right)\partial_{\mu}\phi\partial_{\nu}\phi\mathperiod
\ee
From Eq.~\eqref{h_disformal}, a purely conformal transformation ($D=0$) results in absence of coupling, $h=1$. This is in agreement with the fact that radiation is invariant under conformal transformations.

It is worthwhile noting that although the choice of the disformal transformation matches the one from our initial formulation, the two theories have different physical interpretations. On the one hand, the action \eqref{action} is defined on an ad hoc basis for the specific purpose of the coupling, as commonly done in the literature. On the other hand, in the disformal representation, it is the auxiliary metric $\tilde{g}_{\mu\nu}$ in Eq.\eqref{disformal_final} felt by radiation that indirectly gives rise to the same interaction term. 

This representation can be further understood in analog gravity \cite{barcelo2005}, where the interaction can be modeled using the physical system of electromagnetic waves propagating in the scalar field as a dielectric medium. Let us rewrite the disformal transformation of Eq.~\eqref{disformal_final} by noting that, if $\partial_{\mu}\phi$ is a real time-like vector \cite{Madsen:1988ph}, one can define a scalar-fluid 4-velocity \cite{Ferreira:2018knm,Teixeira:2022sjr,Ferreira:2020fma},
\be\label{4v_scalar}
u_{\mu}=\frac{\partial_{\mu}\phi}{\sqrt{2X}}\mathcomma
\ee
which is a well known fact from the equivalence between a scalar field and a perfect fluid form \cite{Faraoni:2012hn}. Using Eq.~\eqref{4v_scalar} in Eq.~\eqref{disformal_final} and setting $C=1$ without loss of generality, we find the following Gordon metric \cite{Gordon1923}
\be\label{modified_gordon_metric}
\tilde{g}_{\mu\nu} = g_{\mu\nu}+\left( 1-\frac{1}{h^2} \right)u_{\mu}u_{\nu}\mathperiod
\ee
From the perspective of this optical metric, the spacetime with physical metric $g_{\mu\nu}$ can be regarded as containing a flowing scalar fluid of 4-velocity $u_\mu$ and effective refractive index $h$. Accordingly, light rays propagate through the scalar medium along light cones modified by the refractive index, which depends on $\phi$ and its kinetic term $X$.

We now introduce the electric and magnetic susceptibilities of the scalar medium, treated as a linear dielectric. We assume that $\phi$ is homogeneous and evolves only with cosmic time $t$. In Minkowski space, where ${g_{\mu\nu}=\eta_{\mu\nu}}$, the time component of Eq.~\eqref{maxwell2} yields Gauss's law, which can be written in the following form,
\be\label{gauss_law}
\epsilon_0\nabla\cdot\textbf{E}=\rho-\nabla\cdot\left[\epsilon_0\left(h-1\right)\textbf{E}\right]\mathcomma
\ee
where $\mathbf{E}$ is the total electric field, $\rho=j^0/c$ is the free charge density and $\epsilon_0=1/\mu_0c^2$ the permittivity of free space. The last term in the above equation is analogous to a bound charge density, ${\rho_\mathrm{b}= -\nabla\cdot\mathbf{P}}$, introducing the electric polarization of the scalar field, ${\mathbf{P}=\epsilon_0(h-1)\textbf{E}}$, which vanishes in the absence of interaction ($h=1$). According to the definition of the polarization vector, the term ${h-1}$ represents the effective electric susceptibility of the scalar field. On the other hand, the spatial components of Eq.~\eqref{maxwell2}, yield Amp\`{e}re's law,
\be\label{ampere_law}
\nabla \times \mathbf{B}-\mu_0\epsilon_0\frac{\partial\textbf{E}}{\partial t} =  \mu_0\left(\mathbf{J}+\nabla\times\mathbf{M}+\frac{\partial\textbf{P}}{\partial t}\right)\mathcomma
\ee
where $\mathbf{B}$ is the total magnetic field and we have introduced the magnetization ${\mathbf{M}=(1-h)\mathbf{B}/\mu_0}$ which induces the equivalent of a bound current ${\textbf{J}_{\mathrm{b}}=\nabla\times \mathbf{M}}$. Likewise, the temporal change of the electrical polarization of the medium gives rise to an effective polarization current,
\begin{eqnarray}
\textbf{J}_\mathrm{p}&=&\frac{\partial{\mathbf{P}}}{\partial t} \label{polarization_current} \\
&=&\epsilon_0\left(h-1\right)\frac{\partial\textbf{E}}{\partial t}+\epsilon_0\frac{\partial h}{\partial t}\textbf{E}\mathcomma
\end{eqnarray}
where ${\partial h/\partial t=\dot{\phi}(h_\phi+\ddot{\phi}h_X)}$, and the dots represent derivatives with respect to $t$. The velocity of the scalar field contributes to the polarization current, as well as its acceleration due to the dependence on the kinetic term $X$. Using Eqs.~\eqref{gauss_law} and \eqref{ampere_law}, and the constitutive relations, we find the regular macroscopic and inhomogeneous Maxwell equations, valid inside the scalar dielectric:
\begin{align}\label{Maxwell_matter}
\bf{\nabla}\cdot\bf{D} &= \rho \mathcomma \\
\nabla\times\bf{H} &= \bf{J}+\frac{\partial\textbf{D}}{\partial t} \mathcomma
\end{align}
where ${\textbf{D}\equiv\epsilon_0\textbf{E}+\textbf{P}}$ (${=\epsilon_0h\textbf{E}}$) and ${\textbf{H}\equiv \textbf{B}/\mu_0-\textbf{M}}$ (${=h\textbf{B}/\mu_0}$) are the electric displacement and magnetizing fields, respectively. Since ${\mathbf{M}=(-1+1/h)\mathbf{H}}$, its magnetic susceptibility can be identified as ${-1+1/h}$.

In summary, the interacting scalar fluid in which radiation travels appears as a refractive medium in motion, subject to electric and magnetic polarization. The coupling term $h(\phi,X)$ with the electromagnetic sector in the original action \eqref{action} can be modeled as its evolving refractive index, while ${h-1}$ and ${-1+1/h}$ represent its effective electric and magnetic susceptibilities, respectively.

In the next section, we will explore the corresponding cosmological changes in the fine-structure constant by considering that the scalar field, which interacts with the electromagnetic sector, is the quintessence component accelerating the Universe.

\subsection{Cosmological implications}\label{cosmology}

In the total action \eqref{action}, we need to specify the metric $g_{\mu\nu}$ and the matter action $S_m$ to define the cosmological framework. In this regard, we consider the Friedmann-Lema\^itre-Robertson-Walker (FLRW) metric for an homogeneous and isotropic background of a flat universe whose matter content is composed of baryons and cold dark matter. In order to fully define the model at hand, we also specify, in the following, the Lagrangian density of the scalar field dark energy $\mathcal{L}_\phi$, as well as the coupling to the kinetic term of the gauge fields, $h(\phi,X)$, characterizing the interaction between the scalar field and the electromagnetic sector. From now on, we will be working in natural units, where ${c=\hbar=\epsilon_0=\mu_0=1}$.

\subsubsection{The choice for the scalar field dark energy }\label{phi_choice}

We endow the scalar field with dynamics through the choice of the Lagrangian density $\mathcal{L}_{\phi}$. This is germane to the second principle proposed by Bekenstein in Ref.~\cite{Bekenstein:1982eu}. Accordingly, we consider a standard canonical and homogeneous scalar field defined by
\be\label{quintessence_lagrangian}
\mathcal{L}_{\phi}=X-V\mathperiod
\ee
Its dynamics can be found by varying the action Eq.~\eqref{action} with respect to $\phi$. The resulting modified Klein-Gordon equation gives its evolution sourced by the electromagnetic field:
\be
\label{Klein-Gordon}
\square\phi -V_{\phi} = \frac{1}{4}h_{\phi}F_{\alpha\beta}F^{\alpha\beta} + \frac{1}{4}\nabla_{\mu}\left( h_XF_{\alpha\beta}F^{\alpha\beta}\nabla^{\mu}\phi\right)\mathperiod
\ee
It is worthwhile noting the emergence of the last term specific to the $\phi$-kinetic coupling in the above equation.

At this stage, the next step would usually be to assume a specific scalar potential, $V$, to then explicitly find the dynamics of $\phi$. However, one may also work the other way around. That is, to specify a well fundamentally motivated ansatz for the quintessence field and reconstruct the potential {\it a posteriori}. Here, we shall do so by adopting a well known scalar behavior, as a linear function of the number of $e$-folds, $N=\ln a$, where $a$ is the expansion scale factor, such that
\be\label{phi_parametrization}
\kappa\left(\phi - \phi_0\right) = \lambda N\mathcomma
\ee
where $\phi_0$ denotes the present value of the quintessence field, {\it i.e.} ${\phi_0=\phi(N=0)}$, and $\lambda$ is a dimensionless constant. This behavior was already identified in Wetterich's {\it cosmon} model \cite{Wetterich:1994bg}, and the seminal article for conformally coupled quintessence models \cite{Amendola:1999er}. The same parametrization was previously adopted in Ref.~\cite{daFonseca:2021imp} for a quintessence model coupled to dark matter, and in Ref.~\cite{Nunes:2003ff,daFonseca:2022qdf} for a varying $\alpha$ cosmology with ${h\equiv h(\phi)}$. With just one free parameter, $\lambda$, the field evolution conveniently substitutes existing parametrizations of the dark energy equation of state, such as the popular Chevallier-Polarski-Linder (CPL) parametrization \cite{param_a1,param_a2}. Our choice gives rise to the natural behavior of a scalar field during tracking regimes in quintessence models \cite{Copeland:2003cv,Ng:2001hs}, and captures many possible evolutions of dark energy at low redshift, including the cosmological constant when ${\lambda =0}$ \cite{Nunes:2004eog}. In string dilaton models \cite{Gasperini:2001pc}, $\lambda$ acquires different values in each of the cosmological epochs dominated by radiation, matter and dark energy whereas here we assume that it is constant throughout the evolution. In such framework, in the low energy string action of the so called runway-dilaton model, the dilaton can play the role of a quintessence field that drives the late time acceleration, and induces variations in the fine-structure constant\cite{Damour:2002mi,Damour:2010rp}.

This parametrization entirely defines the scalar potential, as we shall see below, by following the procedure of Ref.~\cite{Nunes:2003ff}. In a FLRW background, the energy density for the quintessence field can be written as
\be\label{rho_phi}
\rho_{\phi} = \frac{H^2}{2}\phi'^2 + V\mathcomma
\ee
where $H$ is the Hubble rate and a prime denotes derivative with respect to $N$, {\it i.e.} $\phi'=d\phi/dN$. The dynamics of the scalar field are given by the continuity equation, which can be derived from Eq.~\eqref{Klein-Gordon} with Eq.~\eqref{rho_phi},
\be\label{cons_phi}
\rho_{\phi}'+3H^2\phi'^2=0\mathperiod
\ee
To compute the above equation we have neglected the source terms in the right hand side of Eq.~\eqref{Klein-Gordon}. In a cosmological framework, this is a reasonably good approximation since ${\langle F^{\mu\nu}F_{\mu\nu} \rangle}$ vanishes for a sea of pure radiation \cite{Olive:2001vz}. Additionally, the baryonic contribution is tightly constrained to be small \cite{Leite:2016buh,Leal:2014yqa,Martins:2015dqa} and subdominant with respect to the cosmological Hubble sourced term \cite{Martins:2017yxk}. This assumption has been frequently employed in the literature \cite{Copeland:2003cv,Tavares:2021zhz,daFonseca:2022qdf,vandeBruck:2015rma}. However, there are primordial magnetic fields \cite{Durrer:2013pga} which are argued to be the origin of the present magnetic properties on galactic and extra-galactic scales. Although very weak, their contribution to the right-hand side of Eq.~\eqref{rho_phi} might not be negligible. They are known to have a signature in the cosmic microwave background \cite{Planck:2015zrl} even for nanoscale magnetic fields. It is also possible that dark matter has a magnetic susceptibility \cite{Ramazanov:2020ajq} when interacting with classical electromagnetism. While the inclusion of such an effect is beyond the scope of this work, it is a very interesting challenge for a future analysis.

The Friedmann equation encompassing radiation ($r$), matter ($m$) and the scalar source ($\phi$) reads
\be\label{friedmann}
\frac{H^2}{H_0^2} = \Omega_r^0e^{-4N}+\Omega_m^0e^{-3N}+\frac{\rho_{\phi}}{\rho_0} \mathcomma
\ee
where $\Omega_i^0$ is the present value for the relative energy density of the $i$-species, ${\rho_0=3H_0^2/\kappa^2}$ the present critical energy density, and $H_0$ the current Hubble rate. According to the parametrization of Eq.~\eqref{phi_parametrization}, Eq.~\eqref{cons_phi} can be written as
\be
\rho_{\phi}'+\lambda^2\left(\rho_0\Omega_r^0e^{-4N}+ \rho_0\Omega_m^0e^{-3N}+\rho_{\phi} \right) = 0\mathcomma
\ee
which has the following analytical solution
\be\label{rho_phi_sol}
\frac{\rho_{\phi}}{\rho_0}=\frac{\lambda^2\Omega_r^0}{4-\lambda^2}e^{-4N}+\frac{\lambda^2\Omega_m^0}{3-\lambda^2}e^{-3N}+C_{\phi}e^{-\lambda^2N}\mathcomma
\ee
with the constant
\be
C_{\phi} = 1-\frac{4\Omega_r^0}{4-\lambda^2}-\frac{3\Omega_m^0}{3-\lambda^2}\mathperiod
\ee
Ref.~\cite{daFonseca:2021imp} obtained a similar solution for the energy density of the quintessence field, yet assuming a conformal coupling to dark matter and neglecting radiation. The Friedmann constraint in Eq.~\eqref{friedmann} now reads
\be\label{fried}
\frac{H^2}{H_0^2} = \frac{4\Omega_r^0}{4-\lambda^2}e^{-4N}+\frac{3\Omega_m^0}{3-\lambda^2}e^{-3N}+C_{\phi}e^{-\lambda^2N}\mathperiod
\ee
By combining it with Eqs.~\eqref{rho_phi} and \eqref{rho_phi_sol}, we find the following potential, from which quintessence can arise from a wide range of initial conditions \cite{PhysRevD.61.127301},
\be\label{potential}
V(\phi)= A\,e^{-\frac{4}{\lambda}\kappa\phi}+B\,e^{-\frac{3}{\lambda}\kappa\phi}+C\,e^{-\lambda\kappa\phi}\mathcomma
\ee
where the mass scales are given by
\begin{align}
\label{mass_scales}
A&=\frac{\lambda^2\Omega_r^0}{4-\lambda^2}\frac{H_0^2}{\kappa^2}  \mathcomma  \\
B&=\frac{\lambda^2\Omega_m^0}{3-\lambda^2}\frac{3H_0^2}{2\kappa^2} \mathcomma  \\
C&=\left(1-\frac{\lambda^2}{6}\right)C_\phi \frac{3H_0^2}{\kappa^2}\mathcomma
\end{align}
and where we have fixed, without loss of generality, $\phi_0=0$. The model is invariant under the reflection ${\lambda\mapsto -\lambda}$. This feature stems from the nature of the scalar field parametrization Eq.~\eqref{phi_parametrization}, which has the symmetry ${(\phi,\lambda )\mapsto (-\phi ,-\lambda)}$.

Within this framework, the equation of state for the quintessence field can be written as
\be\label{w_phi}
w_{\phi} = -1+\frac{H^2}{H_0^2}\frac{\rho_0}{\rho_{\phi}}\frac{\lambda^2}{3}\mathperiod
\ee
We are able to put theoretical bounds on $\lambda$ by imposing the acceleration condition on the late time expansion, {\it i.e.} ${w_{\rm eff}^0<-1/3}$, where $w_{\rm eff}$ is the effective equation of state. Evaluating Eq.~\eqref{w_phi} today \cite{daFonseca:2021imp},
\be
w_{\rm eff}^0\approx w_{\phi}^0\Omega_{\phi}^0 = -\Omega_{\phi}^0+\frac{\lambda^2}{3}\mathcomma
\ee
where ${\Omega_{\phi}^0=1-\Omega_m^0-\Omega_r^0}$, one finds the bound
\be
\lambda^2<-1+3\Omega_{\phi}^0<2\mathcomma
\ee
in order for the scalar field to be identified as the quintessence component driving the ongoing acceleration. Naturally, the bound encapsulates smaller values of $\lambda^2$ for decreasing density of dark energy, $\Omega_{\phi}^0$.

\begin{figure*}      \subfloat{\includegraphics[height=0.33\linewidth]{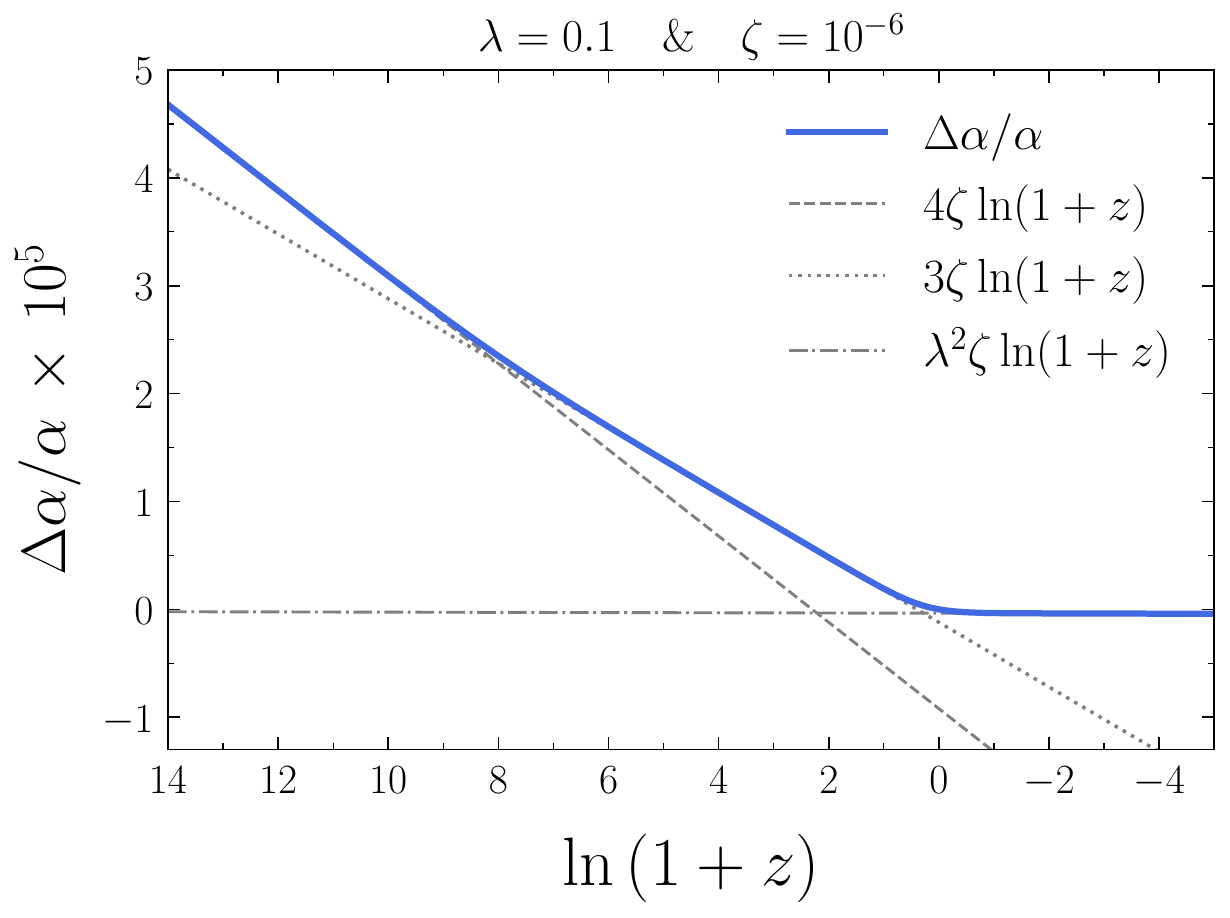}}
      \qquad
      \subfloat{\includegraphics[height=0.33\linewidth]{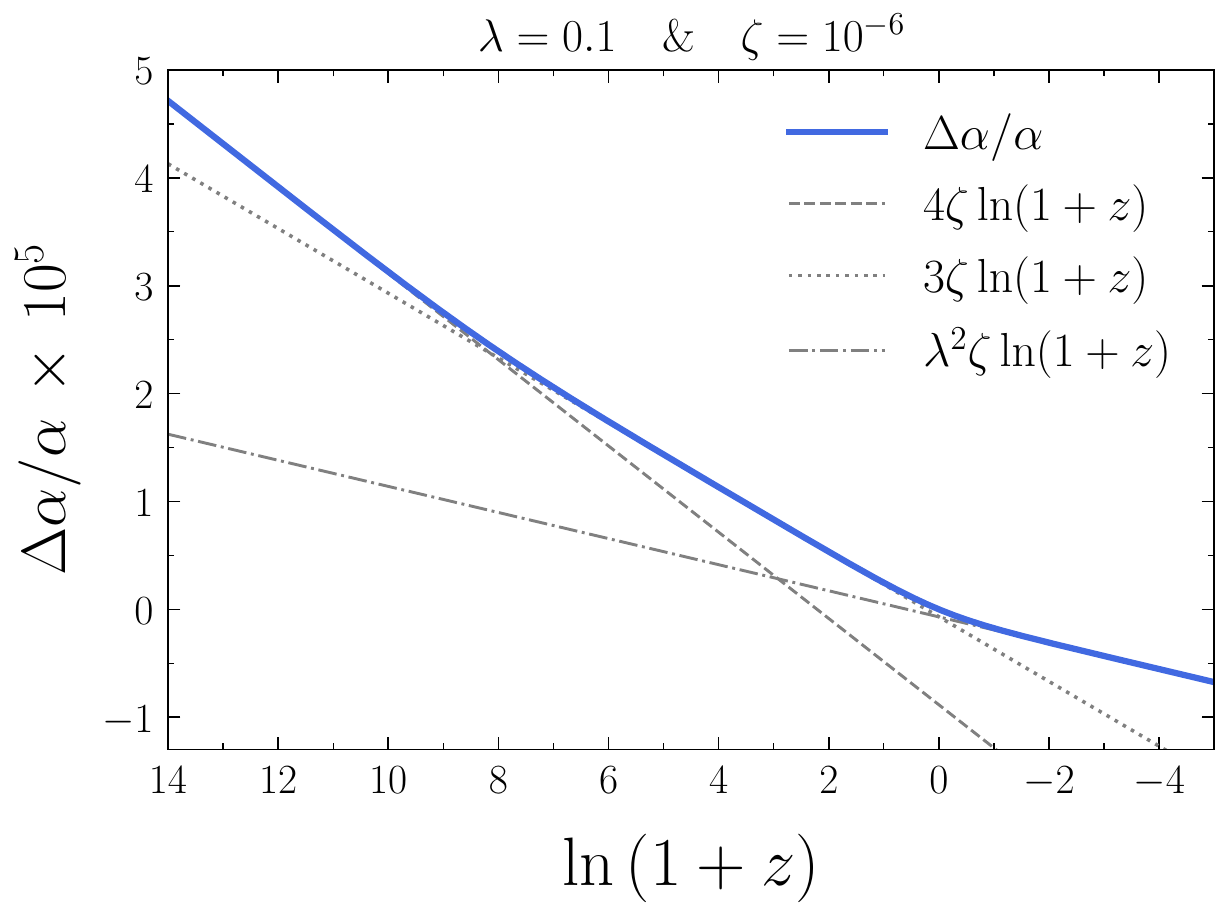}}
  \caption{\label{fig:fig1}Variation of $\alpha$, given by Eq.~\eqref{delta_alpha}, and respective approximations at the cosmological epochs, stated in Eq.~\eqref{behaviour}. It was used the choice of values ${\lambda=0.1}$ (left panel), ${\lambda=1.1}$ (right panel), with ${\zeta = 10^{-6}}$, assuming, without loss of generality, a reference cosmology with ${\Omega_m^0=0.3}$ and  ${\Omega_r^0=10^{-4}}$.}
\end{figure*}

\subsubsection{The choice for the coupling $h(X)$}\label{choice_coupl}

We propose to make the variation of $\alpha$ explicitly depend on the dynamics of dark energy, since the scalar field inducing the variation of the fine-structure constant also provides the accelerating dark energy component. We focus on the specific case where the coupling with the electromagnetic sector in the action \eqref{action} is completely specified by the kinetic energy of the quintessence field: ${h\equiv h(X)}$. 

This specific coupling may emerge through a conformal transformation ${\tilde{g}_{\mu\nu}\mapsto C(X)g_{\mu\nu}}$ \cite{Teixeira:2022sjr,Brax:2020gqg} in the case of non-relativistic fluids, or as a disformal transformation (see Section~\ref{model_intro}). The stability conditions for the kinetically interacting scalar field-matter theories were studied in Ref.~ \cite{Kase:2019veo}. Finally, field derivative couplings were also considered in Ref.~\cite{Pourtsidou:2013nha}, as well as in mimetic models \cite{Shen:2017rya,Vagnozzi:2017ilo} and scalar-fluid models \cite{Boehmer:2015kta,Boehmer:2015sha}, using Brown's formalism \cite{Brown:1992kc}.

To explore the effects of the kinetic coupling on the variation of $\alpha$, we consider explicitly the following power-law function
\be\label{coupling_function}
h(X)=\left(\frac{X_0}{X} \right)^{\zeta}\mathcomma
\ee
for ${X\neq 0}$, where $\zeta$ is the constant strength of the coupling and $X_0$ denotes the present-value of the quintessence kinetic energy, ${X_0=X(N=0)}$. When $X\rightarrow0$, the limit of Eq.~\eqref{coupling_function} is undefined since $X_0$ also vanishes in the absence of quintessence. To remove the ambiguity, we define ${h=1}$ for the exception ${X=X_0=0}$, so that we have a well behaved model, {\it i.e.} recovering non-interacting Maxwell's theory when the field disappears.

A similar power-law form has already been studied to couple quintessence to cold dark matter \cite{Barros:2019rdv,Teixeira:2022sjr} on a phenomenological basis. It leads to viable cosmological models where the current acceleration arises from scaling solutions that alleviate the coincidence problem \cite{delCampo:2008sr}. While the motivation for the interaction mainly stems from cosmological arguments, it is still a phenomenological choice. Therefore, we do not assume any specific fundamental reasons for its emergence. It should be noted, however, that such kinetic couplings occur naturally in low energy actions where matter couples to the scalar field identified as a Goldstone mode of a broken Abelian symmetry \cite{Brax:2016kin}. They are also possible between ultra-light scalars and Standard Model particles or dark matter in effective field theory theories that preserve shift symmetry \cite{Trojanowski:2020xza}. While other operators can be considered in the total action (see \cite{Ramazanov:2020ajq} and references therein), we assume only a general power-law on $X$, in order to isolate its specific effect on the behavior of the model, as is done for the dark matter case of Refs.~\cite{Barros:2019rdv,Teixeira:2022sjr}. In contrast to the latter case, we choose to consider the real range of the coupling strength $\zeta$ in parameter space for the Bayesian inference, including $\zeta>0$. This is to allow for higher values of $\alpha$ in the past, and thus capture both cosmological scenarios. We will see in Sec.~\ref{sec:obs} that the observations seem to favor $\zeta<0$.

Since the strength of the kinetic coupling is expected to be small, the power-law function can be reduced to a logarithm, ${h=1+\zeta\ln(X_0/X)}$. This scaling is line with the running of fundamental couplings with energy \cite{Fritzsch2002}.

Our choice for the parametrization of the scalar field in Eq.~\eqref{phi_parametrization} entails ${2X={\phi^\prime}^2H^2=(\lambda H/\kappa)^2}$ in the FLRW cosmological background. Accordingly, the evolution of $\alpha$ given by Eq.~\eqref{alpha} becomes
\be
\label{alpha2}
\frac{\alpha(N)}{\alpha (0)} = h^{-1} = \left(\frac{H^2}{H_0^2}\right)^{\zeta}\mathperiod
\ee
With $H^2/H_0^2$ given by Eq.~\eqref{fried}, it follows
\be
\label{alpha3}
\frac{\alpha(N)}{\alpha (0)} = \left[\frac{4\Omega_r^0}{4-\lambda^2}e^{-4N}+\frac{3\Omega_m^0}{3-\lambda^2}e^{-3N}+C_{\phi}e^{-\lambda^2N}\right]^\zeta\mathperiod
\ee
Consequently, the temporal variation of $\alpha$ with respect to its present value measured in the lab is quantified by,
 \begin{align}\label{delta_alpha}
 \frac{\Delta\alpha}{\alpha}\equiv&\frac{\alpha (N)-\alpha (0)}{\alpha (0)} = \left(\frac{H^2}{H_0^2}\right)^{\zeta}-1\nonumber \\
 =&\left[\frac{4\Omega_r^0}{4-\lambda^2}e^{-4N}+\frac{3\Omega_m^0}{3-\lambda^2}e^{-3N}+C_{\phi}e^{-\lambda^2N}\right]^\zeta-1 \mathperiod
 \end{align}
The sign of the coupling indicates whether the fine-structure constant was stronger ${(\zeta>0)}$ or weaker ${(\zeta<0)}$ in the past.

Under the premises that the constant $\zeta$ quantifying the deviations is small, we may expand Eq.~\eqref{delta_alpha} to first order
\be\label{first_order}
 \frac{\Delta\alpha}{\alpha}\approx \zeta\ln\left(\frac{H^2}{H_0^2}\right)\mathcomma
 \ee
and find its behavior during the consecutive cosmological epochs dominated by radiation, matter and dark energy (DE):
\be\label{behaviour}
 \frac{\Delta\alpha}{\alpha}\approx
 \begin{cases}
    4\zeta\ln (1+z)+\zeta\ln\left(\frac{4\Omega_r^0}{4-\lambda^2}\right)\mathcomma & \text{radiation}\\[3pt]
    3\zeta\ln (1+z)+\zeta\ln\left(\frac{3\Omega_m^0}{3-\lambda^2}\right)\mathcomma & \text{matter}\\[3pt]
    \lambda^2\zeta\ln (1+z)+\zeta\ln\left(C_\phi\right)\mathcomma & \text{DE}
  \end{cases}
 \ee
where the redshift is defined as ${1+z=a^{-1}=e^{-N}}$.
 
To a good approximation, the variation of the fine-structure constant is therefore a linear function of the number of e-folds whose effective slope has a different value in each epoch. This trend is depicted in Fig.~\ref{fig:fig1} where we plot the values of ${\Delta \alpha / \alpha}$ given by Eq.~\eqref{delta_alpha}, together with its approximations during each cosmological period in Eq.~\eqref{behaviour}. By following the Hubble rate, the variation of $\alpha$ specifically depends on the cosmic content that dominates the Universe at a given epoch. The strongest variation happens during the radiation-dominated era, ${\alpha\propto a^{-4\zeta}}$, until matter takes over, ${\alpha\propto a^{-3\zeta}}$. Afterwards, the $\alpha$ drift slows down dramatically, ${\alpha\propto a^{-\lambda^2\zeta}}$, as the expansion of the Universe accelerates under the domination of dark energy. As ${\lambda\rightarrow 0}$, $\alpha$ freezes to a limiting constant value,
\be
\label{limit}
\frac{\Delta\alpha}{\alpha}\rightarrow\left(\Omega_\phi^0\right)^{\zeta}-1\approx\zeta\ln\Omega_\phi^0\mathcomma
\ee
derived from Eq.~\eqref{delta_alpha}, which depends on the coupling strength and today's cosmic abundances. Therefore, by coupling the dynamics of dark energy to the electromagnetic field, we find that when the scalar field mimics a cosmological constant, it quashes the variations in the fine-structure constant within the late-time acceleration phase. Prior to that time, $\alpha$ continuously evolves in the former radiation and matter dominated eras oblivious to the quintessence component which remains subdominant.

Interestingly, the way $\alpha$ varies presents both disparities and analogies with models where the rate at which it evolves also depends on the cosmological phases \cite{PhysRevD.65.063504, Damour:2002nv} (see also Ref.~\cite{MARTINS2022137002}). In Ref.~\cite{PhysRevD.65.063504} for instance, although the scalar field dictating the variation of $\alpha$ is not quintessence, $\alpha$ remains constant during the radiation-dominated era, instead of varying as in our model. However, in the matter-dominated era, the growth of $\alpha$ is akin to our solution as it depends on the density of matter and is proportional to a logarithm of time. It comes as well to a halt when a cosmological constant accelerates the Universe. As $\alpha$ strongly varies in the radiation-dominated era, it is important to check the compatibility of the kinetic coupling with existing Big Bang Nucleosynthesis (BBN) limits on $\Delta\alpha/\alpha$ (see Section \ref{BBN}).

\section{Parameter constraints}\label{sec:obs}

In this section, we test the kinetically coupled model with different types of observational probes that constrain the variation of the fine-structure constant. They range from measurements of the cosmic microwave background (CMB) in the early Universe, through observations of quasi-stellar objects (QSO) absorption spectra at large redshift, to atomic clocks experiments on Earth. Since we assume that electromagnetism interacts with quintessence, it is also necessary to use specific probes of dark energy. We work with supernovae type Ia (SNIa) standard candles to constrain the cosmological parameters independently of the coupling strength $\zeta$, thus reducing degeneracies between the model parameters. CMB, however, is able to constrain both the cosmology and variations in $\alpha$.

We compute the relevant observables with the Einstein-Boltzmann code CLASS \cite{class}, modified to implement the scalar field potential in Eq.~\eqref{potential} and the evolution of $\alpha$ in Eq.~\eqref{alpha2} which affects the recombination dynamics \cite{Hart:2017ndk}. We carry out likelihood analyses with the inference package MontePython \cite{MP1}. We reconstruct the posterior probability distributions in the parameter space. They correspond to the product of the uncorrelated likelihoods defined below. The parameter space is sampled using either  the Metropolis Hasting or Nested Sampling algorithms \cite{Skilling2006,Feroz_2009,Buchner:2014nha}. The Monte Carlo samples are analysed and plotted with GetDist \cite{Lewis:2019xzd}. The results are reported in Table \ref{tab:constraints} for the posteriors of the coupling strength $\zeta$ in parts per million (ppm) and the quintessence parameter $\lambda$.

\begin{table}[t]
\caption{Constraints on the coupling strength and the quintessence parameter (mean and $68\%$ limits). A SH0ES prior complements the Pantheon sample.}
 \label{tab:constraints}
\centering
\begin{tabular} {l c c}
\hline\hline
Data & $\zeta$ (ppm)  & $\lambda$ \\
\hline
Planck & $-15\pm 240$ & $0.000\pm 0.071$ \\
QSO + Pantheon& $-0.13\pm 0.41$ & $0.00\pm 0.47$ \\
Atomic clocks + Pantheon& $-0.014\pm 0.015$ & $0.00\pm 0.44$ \\
\hline
\end{tabular}
\end{table}

\subsection{Constraints from the CMB in the early Universe}

\begin{figure}[b!]
    \centering
    \includegraphics[scale=0.4]{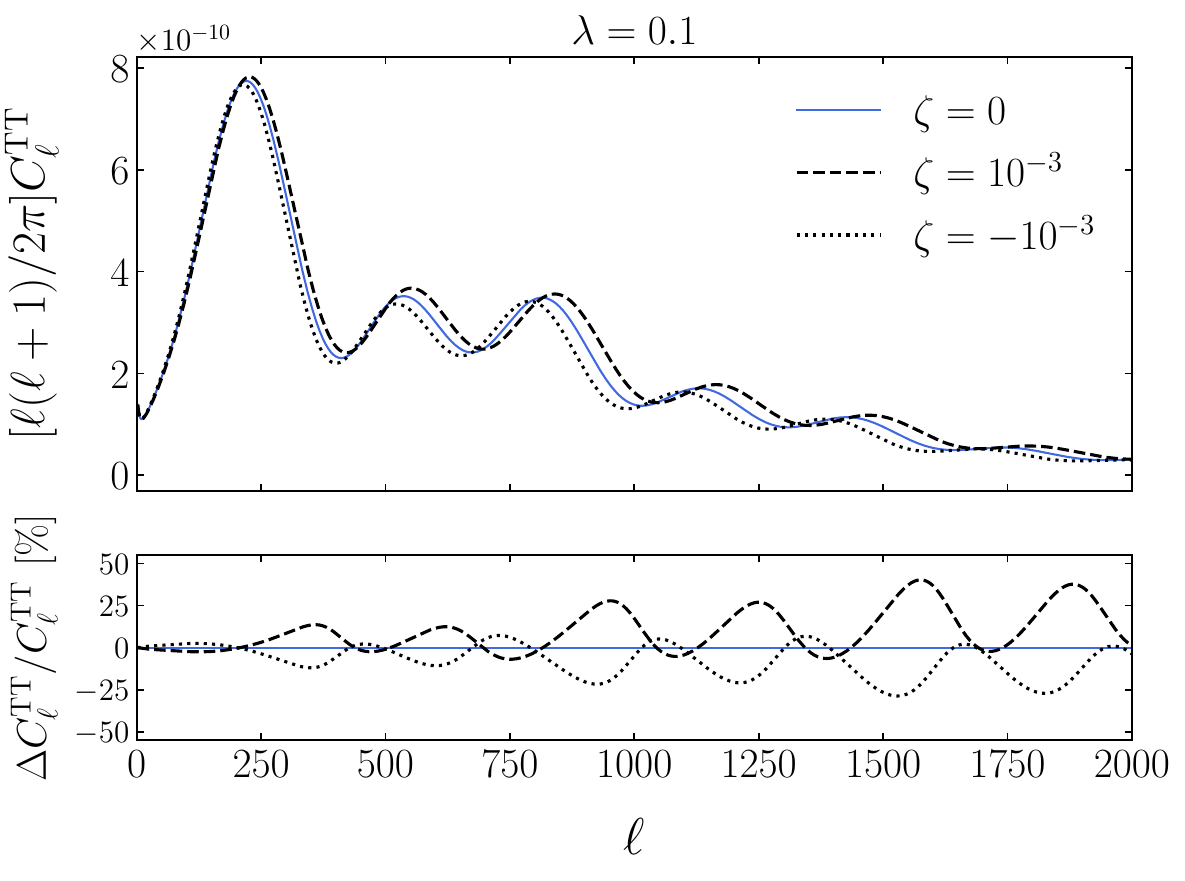}
    \caption{\label{fig_kinetic_CMB} Predicted CMB power spectrum (top panel), for $\lambda=0.1$ and three illustrative values of $\zeta$. Relative differences to the uncoupled case (bottom panel).}
\end{figure}

The Planck measurements of the CMB allow us to constrain the parametrization for dark energy in Eq.\eqref{phi_parametrization} which is governed by the parameter $\lambda$ \cite{daFonseca:2021imp,daFonseca:2022qdf}. On small scales, the amplitude of the first angular acoustic peak increases thanks to a non-negligible dark energy component at decoupling which amplifies the Early-time Integrated Sachs-Wolfe effect. Moreover, the CMB also constrains $\alpha$ at the time of the last scattering \cite{Kaplinghat:1998ry,PhysRevD.60.023515,Battye:2000ds,Hart:2017ndk}. Variations in the fine-structure constant directly affect the ionisation history and the CMB mainly by altering the redshift of recombination, through a change of hydrogen energy levels, and modifying the Thomson scattering cross-section. We show in Fig.~\ref{fig_kinetic_CMB} the effects of the electromagnetic strength $\zeta$ on the temperature anisotropies for a fixed value of $\lambda$. For instance, an increase in $\alpha$ ($\zeta>0$) enhances the peaks amplitude thanks to larger Integrated Sachs-Wolfe effect and suppression of Silk damping on small scales. It also shifts the peaks towards smaller scales as a result of pushing the last scattering surface towards higher redshifts (recombination happens earlier). A thorough explanation of these effects can be found in Sec. 3.6 of \cite{Uzan:2010pm}.

\begin{figure}[t]
    \centering
    \includegraphics[scale=0.55]{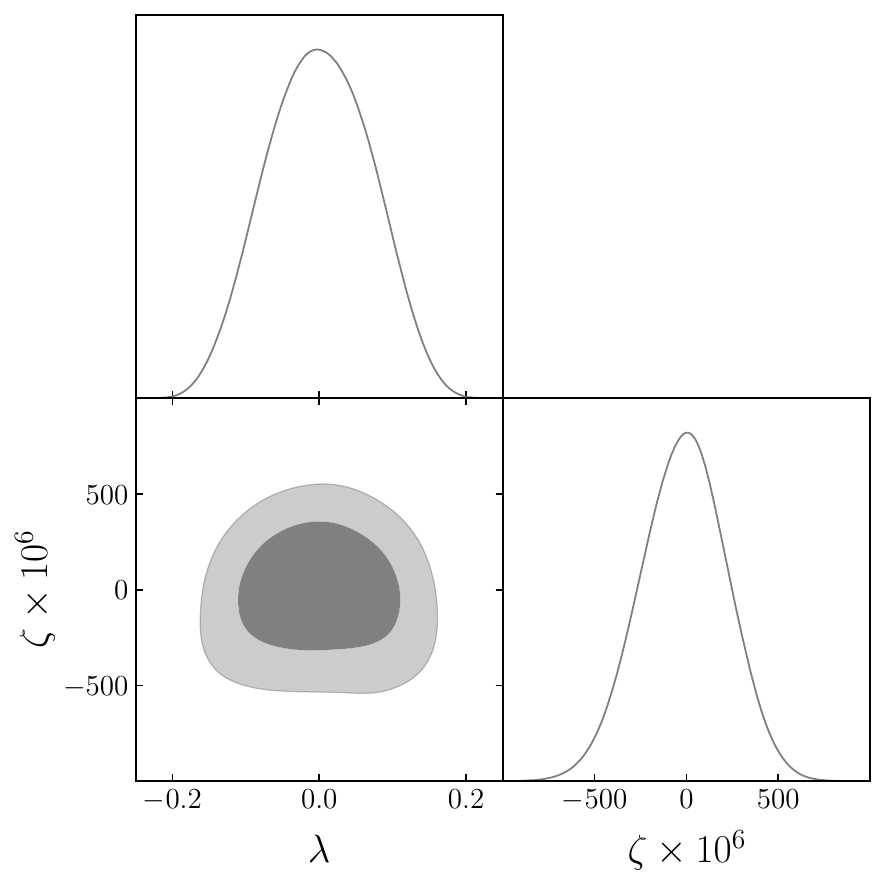}
    \caption{\label{fig_kinetic_planck} Marginalised probability distribution and contours (68\% and 95\% CL) for the Planck analysis.}
\end{figure}

We use the lite version of the high-$l$ Planck likelihood TT,TE,EE \cite{Planck:2019nip}, called ‘Planck’ hereafter, to test our theoretical model and thus constrain the parametrizations of dark energy and $\alpha$. TT, TE and EE denote respectively the angular power spectra of temperature auto-correlation, temperature and polarisation cross-correlation, and polarisation auto-correlation. The free parameters in the Bayesian analysis are ${\{\zeta, \lambda, \omega_c, 100\theta_s, n_s, \ln10^{10}A_s,\tau_\mathrm{reio},A_\mathrm{planck}\}}$, where $\omega_c$ is the physical density of cold dark matter, $\theta_s$ the angular acoustic scale, $n_s$ the scalar spectral index, $A_s$ the amplitude of the primordial power spectrum (normalised at a pivot scale ${k_*=0.05\,\mathrm{Mpc}^{-1}}$), $\tau_\mathrm{reio}$ the optical depth to reionization, and $A_\mathrm{planck}$ the nuisance parameter accounting for the absolute calibration.

The constraint we obtain on the strength of the coupling with the Planck data is
\be
\zeta=\left(-15\pm 240\right)\times 10^{-6}\mathcomma
\ee
at $1\sigma$ confidence level (CL). This result is consistent with recent analysis on the variation of $\alpha$ where parts per thousand constraints were found: ${\Delta\alpha/\alpha=(3.6\pm3.7)\times10^{-3}}$ \cite{Planck:2014ylh} or ${\Delta\alpha/\alpha=(-0.7\pm2.5)\times10^{-3}}$ \cite{Hart:2017ndk} (see also Ref.~\cite{Hart:2019dxi}). At the time of decoupling, around redshift ${z\approx 1100}$, the Universe is near the beginning of matter domination and we can approximate the variation of the fine-structure constant to ${\Delta\alpha/\alpha\sim20\zeta}$ according to Eq.~\eqref{behaviour}. Therefore, the Planck data are indeed expected to constrain $\zeta$ by $\mathcal{O}(10^{-4})$.

Regarding degeneracies with the cosmological parameters, ${(\zeta,\lambda)}$ are poorly correlated as shown in Fig.~\ref{fig_kinetic_planck}. Given that the CMB strongly constrains $\lambda$, close to a cosmological constant, ${\lambda=0.000\pm 0.071}$, $\zeta$ is free to interplay with the other cosmological parameters. This is confirmed by Fig.~\ref{fig_kinetic_planck_ter} of Appendix \ref{appendix_bis}. In the posterior plane ${(H_0,\zeta)}$, the coupling strength is positively correlated to the Hubble constant. As $\zeta$ increases, the first acoustic peak is shifted towards higher multipoles. This shift can be compensated by increased $H_0$ values which reduce the comoving angular diameter distance to the last scattering surface. Likewise, as $\zeta$ increases, the rise in the peaks can be counter-balanced by larger dark matter physical density $\omega_c$. Since the matter-radiation equality takes place earlier, the amplitude of acoustic oscillations are suppressed for a longer period of time. On the contrary, $n_s$ is anti-correlated to $\zeta$. The global slope depends on the tilt of the primordial spectrum and decreasing $n_s$ compensates the extra power on the smallest scales. The same can be achieved by decreasing $A_s$, the global amplitude of the anisotropies power spectra being proportional to that of the primordial spectrum. The opposite effects hold for decreasing $\zeta$ at recombination.

\subsection{Constraints from astrophysical observations of quasars and supernovae}

Since transitions in atoms are directly related to the fundamental constants of nature, the cosmological variation of the fine-structure constant at large look-back-times can be probed with spectroscopic techniques. The Many Multiplet (MM) method \cite{PhysRevA.59.230,1999PhRvL..82..884W} has been extensively used to measure $\Delta\alpha/\alpha$ with the light of distant quasars absorbed by intervening clouds along the line of sight. It consists in comparing shifts in the wavelength of metal transitions \cite{2002PhRvA..66b2501D} produced in quasar absorption systems against laboratory reference values. This method erases cosmological redshift effects by using several atomic species that have transitions showing different dependencies on $\alpha$.

The MM method has led to many direct measurements of $\Delta\alpha/\alpha$ at medium redshift. They can be used to test cosmological models that predict a redshift dependency of $\alpha$, like the one in Eq.~\eqref{behaviour}. The present analysis is carried out with a homogeneous sample (referred to as ‘QSO’) of 26 recent and independent quasar observations with the high-resolution spectrographs VLT/UVES, Keck/HIRES, and Subaru/HDS, as well as the new VLT/ESPRESSO \cite{espresso2021}. These measurements are analogue to those used recently to constrain cosmological models with varying $\alpha$ \cite{Martinelli2021,PhysRevD.105.123507,daFonseca:2022qdf}, thus ensuring a certain degree of comparability among the conclusions of similar studies.

We discard the largest dataset composed of 293 distinct absorption systems from the combination of two samples of archival spectra obtained with UVES and HIRES \cite{PhysRevLett.107.191101,10.1111/j.1365-2966.2012.20852.x,Murphy_2003,https://doi.org/10.48550/arxiv.astro-ph/0310318}. While it suggests a possible dipolar spacial variation of $\alpha$, the measurements suffer from systematic errors related to long-range wavelength distortions caused by the Thorium–argon (ThAr) lamp calibration in the two spectrographs \cite{10.1093/mnras/stt1356}. They could induce spurious shifts in $\Delta\alpha/\alpha$ \cite{10.1093/mnras/stu2420}.

By contrast, the observations we use, which do not support spatial variations \cite{Evans2014,Murphy2016,Kotus2016,MurphyCooksey2017}, are dedicated to test the stability of $\alpha$. The measurements are corrected for the long-range distortions of the wavelength calibration or more resistant to them. Regarding the most recent ESPRESSO measurement \cite{2022espresso}, the wavelength calibration is effectively removed from the error budget, partly thanks to the calibration with a femtosecond-pulsed laser frequency comb (LFC). The data points are listed in Table \ref{tab:QSO} of Appendix \ref{appendix} for the record. More details on the sample can be found in Ref.~\cite{daFonseca:2022qdf}\footnote{We removed the measurement in Ref.~\cite{Agafonova2011} because of possible distortion in the wavelength scale of the UVES spectrograph \cite{Noterdaeme_2021}.} and references therein.

\begin{figure}[b!]
    \centering
    \includegraphics[scale=0.36]{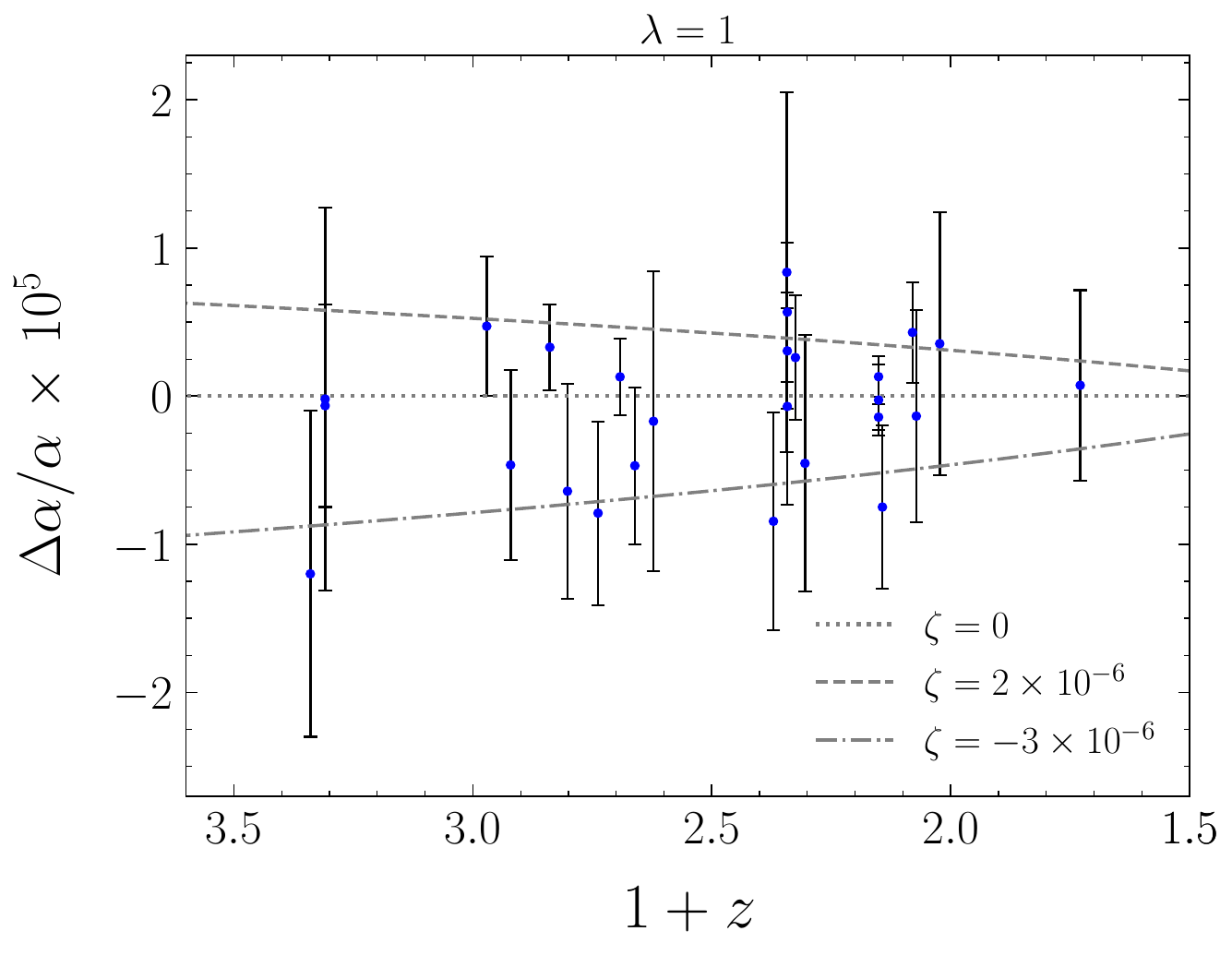}
    \caption{Variation of the fine-structure constant for ${\lambda = 1}$ and different values of $\zeta$. Data points correspond to the 26 measurements of ${\Delta\alpha/\alpha}$ from the QSO sample.}
    \label{fig_qso}
\end{figure}

For an arbitrary value of the scalar field parameter $\lambda$, we illustrate the redshift dependency of $\alpha$ on the coupling strength $\zeta$ in Fig.~\ref{fig_qso} against the 26 well calibrated measurements of $\Delta\alpha/\alpha$. They allow us to test our model in the redshift regime $0.5<z<2.5$. To do so, we define the likelihood of the uncorrelated quasars observations as
\begin{equation}
\label{eq:QSO_likelihood}
\ln \mathcal{L}_{\rm QSO}=-\frac{1}{2}\sum\limits_{i}\frac{1}{\sigma_i^2}\left[\left.\frac{\Delta\alpha}{\alpha}\left(z_i\right)\right|_\textrm{\tiny{th}}-\left.\frac{\Delta\alpha}{\alpha}\left(z_i\right)\right|_\textrm{\tiny{obs}}\,\right]^2\mathcomma
\end{equation}
where the subscripts `th' and `obs' denote the model prediction in Eq.~\eqref{delta_alpha} and the $i^{\textrm{\tiny{th}}}$ data point, respectively. $\sigma_i$ is the measurement error and $z_i$ is the system absorption redshift of a given data point.

\begin{figure}[b]
    \centering
    \includegraphics[scale=0.55]{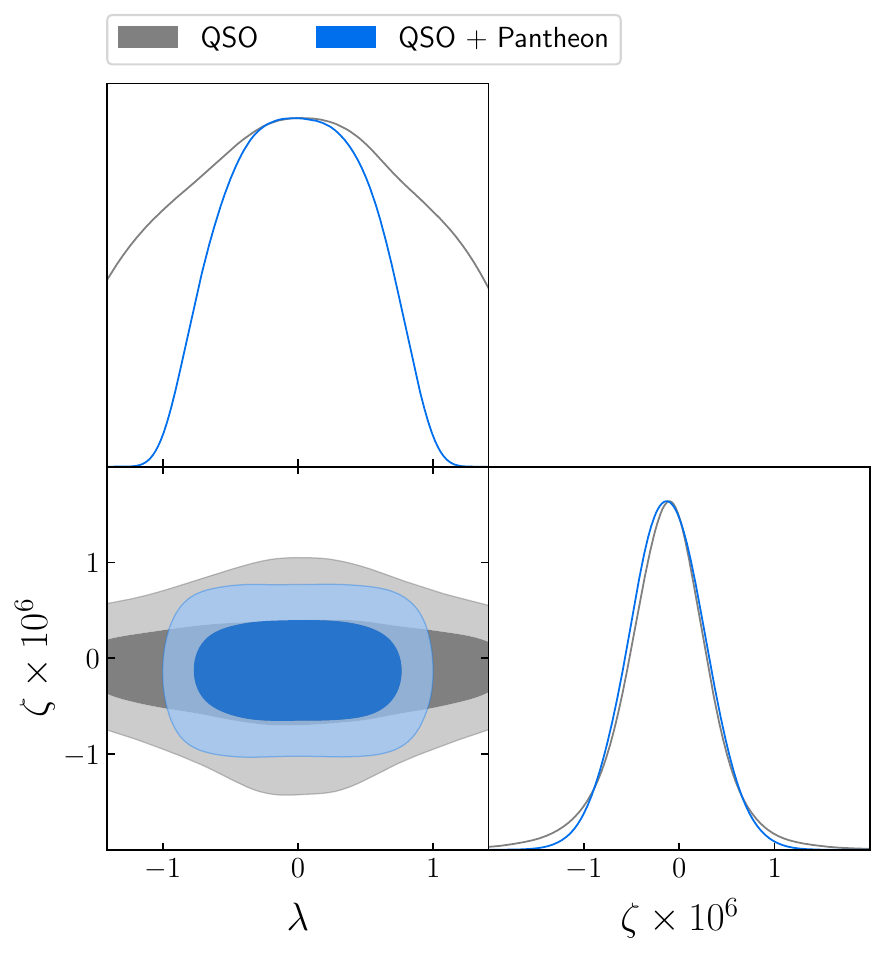}
    \caption{\label{fig_kinetic_qso} Marginalised probability distribution and contours (68\% and 95\% CL) for the QSO analysis, with and without the Pantheon sample (including the SH0ES prior).}
\end{figure}

Furthermore, to reduce their degeneracies with $\zeta$, we specifically constrain the cosmological parameters ${\{\lambda,\Omega_c,H_0\}}$, in a similar redshift regime, with SNIa observations and a SH0ES prior which are insensitive to ${\Delta\alpha/\alpha}$. $\Omega_c$ is today's cold dark matter fractional energy density. As a dark energy probe, SNIa are widely popular to test cosmological models at the background level \cite{10.1111/j.1365-2966.2011.19584.x}. Being standard candles, they can probe the luminosity distance as a function of redshift. Usually, this is done with the distance modulus $\mu$,
\begin{equation}
\mu=m-M=5\log (d_L)+25\mathcomma
\label{eq:distance_modulus}
\end{equation}
where $m$ is the supernova apparent magnitude, $M$ is the absolute magnitude fitted as a free parameter in the analysis, and $d_L$ is the luminosity distance in Megaparsec defined as
\begin{equation}
d_L(z)=\left(1+z\right)\int_0^z\frac{d\bar{z}}{H(\bar{z})}\mathcomma
\label{eq:dL_lambda}
\end{equation}
where $H$ is given by Eq.~\eqref{fried}. Since $\lambda$ alters comoving distances, through the Hubble expansion Eq.~\eqref{fried}, it can be constrained by SNIa distance measurements.

We use the Pantheon sample \cite{pantheon} which is the largest compilation of SNIa to date. It contains 1048 data points within the redshift interval $0.01<z<2.3$, consisting of the supernovae apparent magnitude $m(z)$ and the respective measurement uncertainties. In the Bayesian analysis, we compute the luminosity distance with CLASS to predict the theoretical distance modulus in Eq.~\eqref{eq:distance_modulus}. We fit the model to the Pantheon sample with the following likelihood \cite{Betoule:2014frx},
\begin{equation}
\ln \mathcal{L}_{\rm SN}=-\frac{1}{2}\left(\mu_\mathrm{obs}-\mu_\mathrm{th}\right)^{\rm T}C^{-1}\left(\mu_\mathrm{obs}-\mu_\mathrm{th}\right),
\end{equation}
where the vectors $\mu_\mathrm{obs}$ and $\mu_\mathrm{th}$ contain the brightness measurements and the model predictions for the different redshifts, respectively, and $C$ is the covariance matrix. We also put a Gaussian prior on the Hubble constant $H_0$ from the SH0ES collaboration, updated to Ref.~\cite{Riess_2019}. The present value of the expansion rate of the Universe was estimated by Hubble Space Telescope observations of 70 long-period Cepheids in the Large Magellanic Cloud. Finally, we fix $\Omega_b=0.05$ in the analysis. Overall, when the Bayesian inference includes the QSO and Pantheon observational samples, and the SH0ES prior, the free parameters are ${\{\zeta,\lambda,\Omega_c,H_0,M\}}$, where the absolute magnitude, $M$, is a nuisance parameter kept free in the inference.

The results for the quasars combined with the supernovae likelihood analysis (including the Cepheids prior) are reported in Fig.~\ref{fig_kinetic_qso}, where the marginalised contours (68\% and 95\% confidence levels) are presented. Comparatively to the Planck results, we obtain an increase of precision of three orders of magnitude for the coupling strength, with a mean value of 
\be
\zeta=\left(-0.13\pm 0.41\right)\times 10^{-6}\mathperiod
\ee
Within the QSO sample, the measurement of Ref.~\cite{Kotus2016}, $\Delta\alpha/\alpha=-1.42\pm 0.85$ ppm at redshift $z=1.15$, is the data point with the highest statistical weight (see Table \ref{tab:QSO} of Appendix \ref{appendix}). Using Eq.~\eqref{behaviour} in the matter-dominated era, one can confirm that $\zeta$ may indeed be constrained by $\mathcal{O}(10^{-7})$. We see that the addition of the Pantheon sample, together with the SH0ES prior, marginally improves the constraint on $\zeta$, mostly at $2\sigma$ CL. Similarly to the CMB prediction, the mean of $\zeta$ slightly deviates towards a negative value. This result is consistent with previous studies, such as in Ref.~\cite{PhysRevD.105.123507}, although there the gauge kinetic takes a different form, being a linear function of the scalar field, $h\propto\zeta\phi$.

Regarding the quintessence parameter, we still find a mean value centered at zero, however, with looser constraints than those from Planck: $\lambda=0.00\pm 0.47$. Since quasar experiments are barely sensitive to $\lambda$, the constraints on this parameter mainly stem from the supernovae observations, which are able to probe dark energy signatures.

\subsection{Constraints from atomic clocks experiments on Earth}

Observing the long-time demeanour of periodic transitions in different atoms puts constraints on the drift rate of the fine-structure constant \cite{2008Sci...319.1808R}. This is done by measuring frequencies of different transitions in atomic clocks that adopt different atoms throughout a period of 2-4 yrs (see Section 3.1 of \cite{Uzan:2010pm}). For example, in the hydrogen atom, the frequency associated with the fine-structure is $\nu \propto cR_H\alpha^2$, where $R_H$ is the Reydberg constant. Thus by measuring the variations on the frequency throughout a long period of time, $\dot{\nu}$, we are able to place bounds on the time variation of $\alpha$. For our present model, the $\alpha$ drift rate is
\be
\frac{\dot{\alpha}}{\alpha} = 2\zeta H'\mathcomma
\ee
which evaluated at present times yields
 \be
 \label{drift}
 \left.\frac{\dot{\alpha}}{\alpha}\right|_0 = -\zeta H_0\left(\lambda^2+3\Omega_m^0+4\Omega_r^0\right)\mathperiod
 \ee

The strongest limit on the current drift rate of the fine-structure constant obtained from atomic clock comparisons in Earth laboratories is currently the following \cite{2021PhRvL.126a1102L}:
\be
 \left.\frac{\dot{\alpha}}{\alpha}\right|_0=\left(1.0\pm1.1\right)\times 10^{-18}\,\,\textrm{yr}^{-1}\mathperiod
\label{drift_rate_bound}
\ee

We notice that as the field approaches a cosmological constant, ${\lambda\rightarrow 0}$, higher absolute values of $\zeta$ are allowed such that the drift rate remains within the bounds placed by Eq.~\eqref{drift_rate_bound}. We also notice that the bound favors negative values of $\zeta$.

We define the corresponding likelihood, referred to as `atomic clocks', as
\be
\label{clocks_likelihood}
\ln\mathcal{L}_\textrm{\tiny{clocks}}=-\frac{1}{2}\frac{\left[\left.\left(\frac{\dot{\alpha}}{\alpha}\right)_0\right|_\textrm{\tiny{th}}-\left.\left(\frac{\dot{\alpha}}{\alpha}\right)_0\right|_\textrm{\tiny{obs}}\right]^2}{\sigma^2}\mathcomma
\ee
where the subscript `th' stands for the theoretical prediction of the model in Eq.~\eqref{drift} and `obs' the observed bound in Eq.~\eqref{drift_rate_bound}. The measurement error is denoted $\sigma$.
\begin{figure}[t]
    \centering
    \includegraphics[scale=0.55]{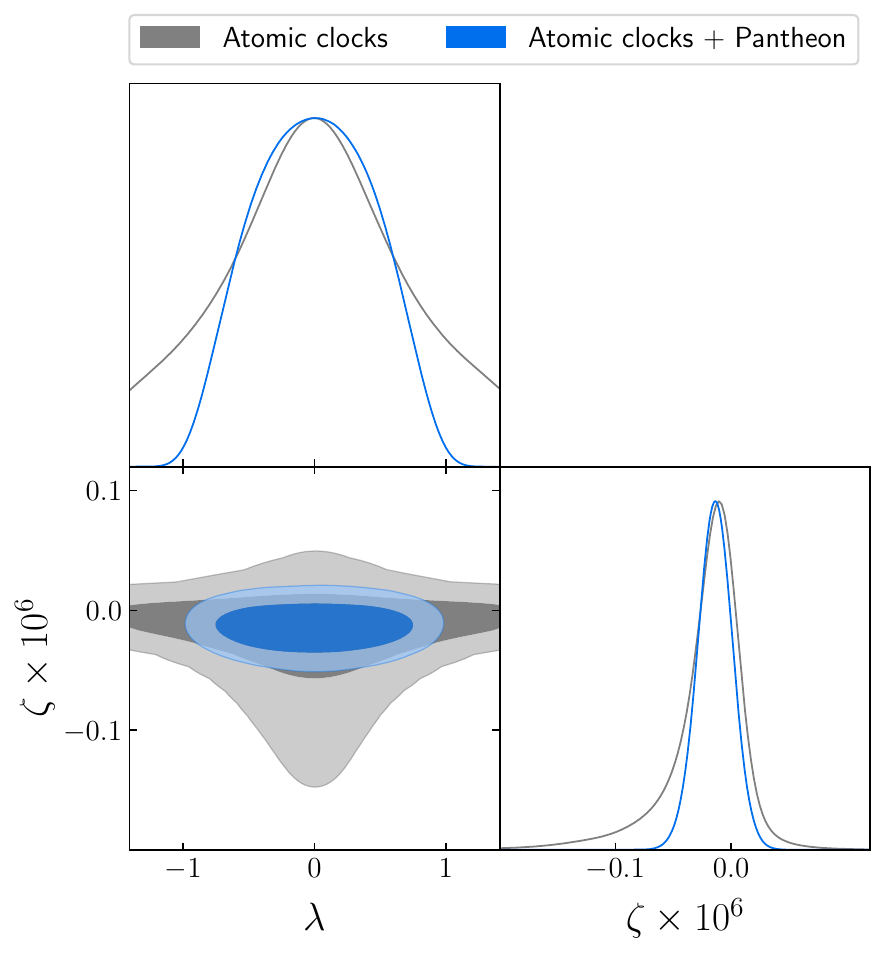}
    \caption{\label{fig_kinetic_clocks} Marginalised probability distribution and contours (68\% and 95\% CL) for the atomic clocks analysis, with and without the Pantheon sample (including the SH0ES prior).}
\end{figure}
To break the degeneracies from Eq.~\eqref{drift} between the coupling strength and the cosmological parameters, we again use the Pantheon sample and the SH0ES prior, being observations that do not probe ${\Delta\alpha/\alpha}$. The free parameters are the same as in the QSO analysis.

As a result of the Bayesian analysis, the local measurement of the $\alpha$ drift rate tightens the coupling bound to 
\be
\zeta=\left(-0.014\pm 0.015\right)\times 10^{-6}\mathcomma
\ee
which represents an improvement in constraining power by one order of magnitude compared to the quasars observations, and a four order of magnitude improvement with respect to the CMB data. According to Eqs.~\eqref{drift} and \eqref{drift_rate_bound}, one can indeed expect a constraining power on $\zeta$ of the order of $\mathcal{O}(10^{-8})$ since ${H_0\approx 0.7\times 10^{-10}}$ $\rm yr^{-1}$. We can see in Fig.~\ref{fig_kinetic_clocks} that the supernovae observations placing stringent bounds on $\lambda$ tightens the constraints on $\zeta$ by compressing the confidence contours in the $(\lambda,\zeta)$ plane. This can also be verified through Eq.~\eqref{drift_rate_bound}. As $\lambda$ decreases, the bounds in $\zeta$ slightly narrow as well. Being specifically constrained by the Pantheon sample, the scalar field is obviously still consistent with a cosmological constant, $\lambda=0.00\pm 0.44$, like in the previous subsection. It is worth noting that, in contrast to usual models \cite{PhysRevD.89.083509,Martinelli2021,PhysRevD.105.123507,daFonseca:2022qdf}, tighter constraints on the dark energy parameter does not weaken the constraints on the electromagnetic coupling.
 
\subsection{WEP and BBN tests}
\label{BBN}

We now aim to confirm whether the constraints obtained through the previous statistical treatment are compatible with both the weak equivalence principle, whose experimental precision is particularly high at the local level, and the existing constraints derived from BBN on the variation of $\alpha$ at very high redshift.

Whereas the WEP states that two bodies of different mass or composition falling in a gravitational field are equally accelerated, any theory with dynamical fundamental constants causes a violation of the universality of free fall. A coupling to classical electromagnetism induces a fifth force, in our present framework caused by the scalar degree of freedom \cite{Damour:2010rp}. The composition of an object naturally depends on the mass of the elementary particles and also possesses a contribution from the binding energy of all the interactions. The total mass of a body is hence a function of the fundamental constants $\alpha_i$. Any variation in $\alpha_i$ brings about an extra force to the equations of motion \cite{Uzan:2010pm}:
\be
\vec{a} = -c^2\frac{\partial m}{\partial\alpha_i}\vec{\nabla} \alpha_i\mathcomma
\ee
where $m$ is the mass of the object. This effect depends on the chemical composition of the given object, or, more precisely, how the material making it is sensitive to the fundamental constant $\alpha_i$.

As the universality of free fall underpins General Relativity, we put our kinetic model with varying $\alpha$ to the test, by making a comparison between the constraints obtained on $\zeta$ and the level of the WEP violation quantified by the E\"otv\"os ratio, $\eta$. We take advantage of the following model-dependent relation \cite{Carroll:1998zi,Dvali:2001dd,Chiba:2001er},
\be
\eta\approx 10^{-3}\zeta^2 \mathperiod
\ee
This relation is based on a linear dependency on $\phi$ for the evolution of $\alpha$, expanding it around its present value $\alpha_0$, which can always be done. Since our choice for $h(X)$ also leads to a similar dependency, by expanding the last exponential term in Eq.\eqref{alpha3}, we can safely apply it to the present model.

By having two different bodies made of dissimilar compositions on a circular orbit around our planet, the WEP can be tested by trying to detect any additional (fifth) force, besides gravity, causing the bodies to deviate from its maintained circular orbit. This is precisely the objective of the MICROSCOPE mission \cite{Berg__2015}, where the bodies are alloys made out of titanium and platinum. It has obtained the most stringent bound on $\eta$ to date \cite{PhysRevLett.129.121102}:
\be
\label{eta}
\eta=\left(-1.5\pm 2.8\right)\times 10^{-15}\mathperiod
\ee

From the above equation, local tests of the equivalence principle limit the magnitude of the coupling parameter to 
\be
|\zeta|<1.14\times 10^{-6}\mathperiod
\ee
We reach the conclusion that the results from both the quasars absorption lines, ${\zeta=-0.13\pm 0.41}$ ppm, and the atomic clocks experiments, ${\zeta=-0.014\pm 0.015}$ ppm, pass the WEP test. Conversely, the CMB constraint, ${\zeta=-15\pm 240}$ ppm, falls short by two order of magnitude.

As far as the BBN is concerned, it is the earliest probe of the variation of the fine-structure constant. Deviations in $\alpha$ mainly alter the production of primordial nuclei by changing the mass difference between proton and neutron, ${Q=m_n-m_p=1.30\,\rm MeV}$, as well as the Coulomb barrier affecting the nuclear reaction rates \cite{IOCCO20091}. The primordial ratio of neutrons to protons, $n/p$, at neutron freeze-out dictates the abundance of Helium-4 since most of the neutrons are processed into it during nucleosynthesis. The Helium-4 mass fraction is thus mostly sensitive to
\be
\frac{n}{p}\sim\exp\left(-\frac{Q}{T}\right)\mathcomma
\ee
where ${T\sim0.8}$ MeV is the freeze-out temperature. For example, larger mass differences decrease the $n/p$ ratio at freeze-out, decreasing in turn the final helium abundance. The change in $Q$ can be parameterized as a function of ${\Delta\alpha/\alpha}$ based on the electromagnetic quark masses and binding energy determined by the value of $\alpha$ \cite{GASSER198277}. The other nuclei are produced through different nuclear reactions where tunnelling plays a crucial role, particularly in the case of Lithium-7 production.

Since the present model predicts that $\alpha$ continuously varies during the radiation-dominated era, we have to test our model against the limits obtained from BBN at redshift ${z\sim 4\times 10^8}$. We use the strong ppm constraint recently derived in the context of Grand Unified Theories from Deuterium and Helium-4 abundances \cite{refId0}:
\be
\frac{\Delta\alpha}{\alpha}=\left(2.1^{+2.7}_{-0.9}\right)\times 10^{-6}\mathperiod
\ee

At that time, Eq.~\eqref{behaviour} gives ${\Delta\alpha/\alpha\sim 70\zeta}$. The BBN thus provides a limit on the coupling parameter, 
\be
\zeta\sim \mathcal{O}\left(10^{-7}\right)\mathcomma
\ee
at very high redshift, in agreement with the atomic clocks bound measured locally. While the QSO constraint on $\zeta$ is competitive with the BBN limits, Planck is weaker by three orders of magnitude.

\section{Conclusions}\label{sec:conclusions}

Until now, varying $\alpha$ models have been assuming a multiplicative field dependent function, $h(\phi)$, in the electromagnetic Lagrangian. However, in this work, we explored an original model driven by a scalar kinetic interaction, $h(X)$, to Maxwell's theory. Since the scalar field also provides the dark energy component, its dynamics is expected to play a coincident role in the interaction and in the late-time acceleration, weaving an explicit link between the two. Here, the evolution of the gauge field $A_\mu$, given by Eq.~\eqref{maxwell2}, is sensitive to the kinetic term of the quintessence field. The phenomenology of such a framework can arise from a disformal transformation in which photons follow geodesics of null length of an effective Gordon metric related to $g_{\mu\nu}$. We highlighted how this treatment leads to interpreting the interacting quintessence as an optical medium in motion with an effective refractive index $h$ that modifies the light cones along which radiation propagates.

The field equations were evolved in a flat FLRW background with a simple parametrization for the scalar degree of freedom, anchored in fundamental theories and motivated by a well established dark energy phenomenology. At the background level, the quintessence component is fully parameterized by one parameter, ${\lambda = \kappa\phi^\prime}$, limiting the degeneracies in the statistical analysis. We derived the scalar energy density and potential which entirely characterize the quintessence behavior in a universe filled with radiation, pressureless matter and dark energy. As for the choice of the kinetic coupling, we relied on a power law form, ${X^{\zeta}}$, which has already proven itself in the study of couplings within the dark sector.

We have found that the variation of the fine-structure constant can be approximated to a linear function of ${\ln(1+z)}$, with an effective slope specified by the successive dominant energy sources. $\alpha$ evolves differently during the radiation and matter epochs, and freezes in the limit of a cosmological constant throughout the dark energy dominated era. This remarkable feature makes sense bearing in mind that the coupling is naturally expressed in terms of the Hubble rate, as seen in Eq.~\eqref{alpha2}.

The model was fitted to observations that constrain ${\Delta\alpha/\alpha}$ within a wide redshift range: from the CMB to local atomic clocks experiments, through absorption lines of quasars light. We combined them with SNIa and Cepheids data when required to specifically constrain the cosmological parameters in consistent redshift regimes, and reduce their possible degeneracies with the coupling parameter. The Bayesian analysis was carried out with a modified version of the CLASS code to compute the cosmological evolution and its observables, together with the MontePython code to sample the parameter space and produce the Monte Carlo chains. We then compared the constraints obtained on $\zeta$ to those expected from local WEP tests, and from the BBN as the earliest probe of ${\Delta\alpha/\alpha}$.

Regarding the strength of the coupling, the Planck data weakly constrain it while the atomic clocks bound gives the most stringent constraint, ${\zeta=-0.014\pm 0.015}$ ppm. It improves that of Ref.~\cite{PhysRevD.105.123507} by a factor of ten, although the gauge kinetic function and dataset combination are different. The QSO sample provides a great improvement, ${\zeta=-0.13\pm 0.41}$ ppm, compared to the 100 ppm precision obtained in Ref.~\cite{daFonseca:2022qdf}. Whereas in the latter reference it is the perfectly degenerate product $\lambda\zeta$ which is being constrained, in the present work $\zeta$ and $\lambda$ are mostly decorrelated from each other outside dark energy domination (see Eq.\eqref{behaviour}). This is also in contrast to other common parametrizations where the dark energy equation of state parameter is strongly degenerate with the coupling \cite{PhysRevD.89.083509,Martinelli2021,PhysRevD.105.123507}. Interestingly, the QSO result does not violate locally the universality of free fall, and is competitive with the limit imposed by primordial nucleosynthesis. The constraint from the atomic clocks experiment for its part is in line with WEP and BBN, while the constraint from the Planck data is, as expected, too weak with respect to both.

The gauge kinetic function, $h$, can eventually be written as a dependency on $\phi$ rather than $X$ through our choice of the scalar field parametrization, given by Eq.~\eqref{phi_parametrization} in an FLRW background. It is worthwhile noting, however, that this resulting dependency differs from the familiar Taylor expansion at first order, $h\propto\zeta\phi$, and produces a noteworthy phenomenology whereby the behavior of $\alpha$ is controlled by the consecutive cosmological phases. In fact, we studied and tested a motivated departure from the linear form of the gauge kinetic function. We have therefore devised a novel form of the coupling between the two sectors, founded on the dynamics of dark energy, which may serve as an archetype for future research.


\acknowledgments
The authors thank Tiago Barreiro and Nelson Nunes for their invaluable comments. We also thank the anonymous referee for the very thoughtful suggestions that helped improve the quality of the manuscript B.J.B. is supported by the South African NRF Grants No. 120390, reference: BSFP190416431035; No. 120396, reference: CSRP190405427545. V.d.F. acknowledges FCT support under the grant reference 2022.14431.BD. This work was financed by FEDER -- Fundo Europeu de Desenvolvimento Regional -- funds through the COMPETE 2020 -- Operational Programme for Competitiveness and Internationalisation (POCI) -- and by Portuguese funds of FCT -- Funda\c c\~ao para a Ci\^encia e a Tecnologia -- under projects PTDC/FIS-AST/0054/2021 and UIDB/04434/2020 \& UIDP/04434/2020.


\appendix
\section{QSO sample}\label{appendix}

We list in the table below the measurements of the QSO dataset used in the statistical analysis. 

\begin{table}[h!]
\caption{QSO sample. Listed are, respectively, the name of the source, the absorber redshift, the measurement of $\Delta\alpha/\alpha$ in ppm and its error at $1\sigma$ CL, and the original reference.}
\label{tab:QSO}
\begin{tabular} {c c c c}
\hline\hline
Quasar &$\quad z_\textrm{abs}\quad$ & $\quad\Delta\alpha/\alpha$ (ppm)$\quad$ & Reference\\
\hline
J0120{\ensuremath{+}}2133 & $0.729$ & $0.73\pm6.42$ & \cite{MurphyCooksey2017} \\
J0026{\ensuremath{-}}2857& $1.023$ & $3.54\pm8.87$ & \cite{Murphy2016} \\
J0058{\ensuremath{+}}0041 & $1.072$ & $-1.35\pm7.16$ & \cite{Murphy2016} \\
3 quasars & $1.080$ & $4.30\pm3.40$ & \cite{SongailaCowie2014} \\
HS1549{\ensuremath{-}}1919 & $1.143$ & $-7.49\pm5.53$  & \cite{Evans2014} \\
HE0515{\ensuremath{-}}4414 & $1.151$ & $1.31\pm1.36$ & \cite{2022espresso} \\
HE0515{\ensuremath{-}}4414 & $1.151$ & $-1.42\pm0.85$ & \cite{Kotus2016} \\
HE0515{\ensuremath{-}}4414 & $1.151$ & $-0.27\pm2.41$ & \cite{Milakovic2020} \\
J1237{\ensuremath{+}}0106 & $1.305$ & $-4.54\pm8.67$ & \cite{Murphy2016} \\
J0120{\ensuremath{+}}2133 & $1.325$ & $2.60\pm4.19$ & \cite{MurphyCooksey2017} \\
HS1549{\ensuremath{+}}1919 & $1.342$ & $-0.70\pm6.61$ & \cite{Evans2014} \\
J0841{\ensuremath{+}}0312 & $1.342$ & $3.05\pm3.93$ & \cite{Murphy2016} \\
J0841{\ensuremath{+}}0312 & $1.342$ & $5.67\pm4.71$ & \cite{Murphy2016} \\
 J0120{\ensuremath{+}}213 & $1.343$ & $8.36\pm12.16$ & \cite{MurphyCooksey2017} \\
J0108{\ensuremath{-}}0037& $1.371$ & $-8.45\pm7.34$  & \cite{Murphy2016} \\
J1029{\ensuremath{+}}1039& $1.622$ & $-1.70\pm10.11$  & \cite{Murphy2016} \\
HE1104{\ensuremath{-}}1805& $1.661$ & $-4.70\pm5.30$ & \cite{SongailaCowie2014} \\
HE2217{\ensuremath{-}}2818 & $1.692$ & $1.30\pm2.60$  & \cite{molaro2013} \\
HS1946{\ensuremath{+}}7658 & $1.738$ & $-7.90\pm6.20$  & \cite{SongailaCowie2014} \\
HS1549{\ensuremath{+}}1919 & $1.802$ & $-6.42\pm7.25$ &  \cite{Evans2014} \\
Q1103{\ensuremath{-}}2645& $1.839$ & $3.30\pm2.90$  & \cite{BainbridgeWebb2017} \\
Q2206{\ensuremath{-}}1958 & $1.921$ & $-4.65\pm6.41$  & \cite{Murphy2016} \\
Q1755{\ensuremath{+}}57 & $1.971$ & $4.72\pm4.71$  & \cite{Murphy2016} \\
PHL957 & $2.309$ & $-0.65\pm6.84$  & \cite{Murphy2016} \\
PHL957 & $2.309$ & $-0.20\pm12.93$  & \cite{Murphy2016} \\
J0035{\ensuremath{-}}0918 & $2.340$ & $-12.0\pm11.0$  & \cite{Welsh2020} \\
\hline
\end{tabular}
\end{table}

\section{MCMC chain plots}\label{appendix_bis}

\begin{figure*}[t]
    \centering
    \includegraphics[scale=0.43]{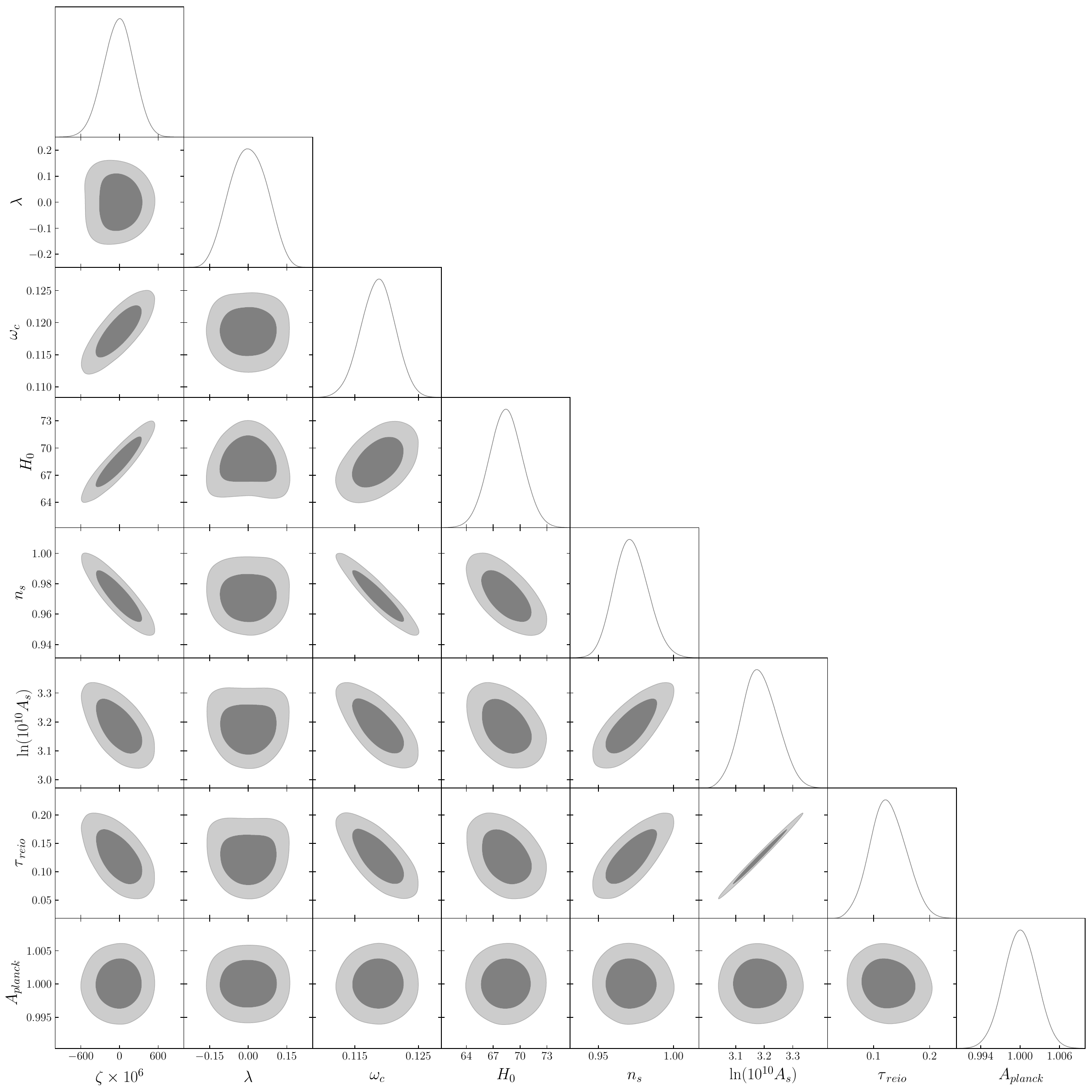}
    \caption{Marginalised contours and posterior distributions (68\% and 95\% confidence levels) of the Planck analysis for the coupling parameter $\zeta$, the cosmological parameters and the nuisance parameter $A_{\rm planck}$.}
    \label{fig_kinetic_planck_ter}
\end{figure*}

\begin{figure*}[t]
    \centering
    \includegraphics[scale=0.44]{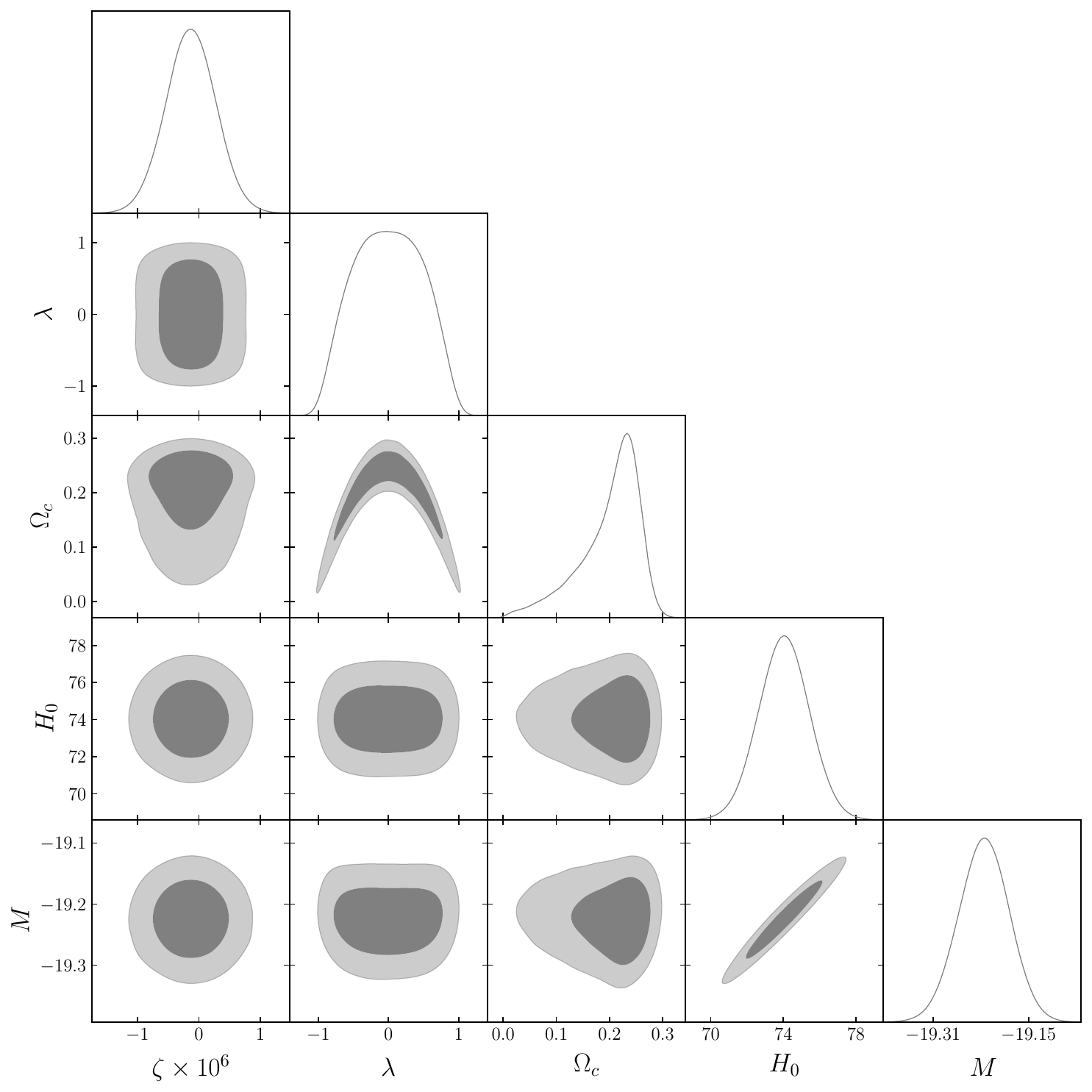}
    \caption{Marginalised contours and posterior distributions (68\% and 95\% confidence levels) of the QSO + Pantheon + SH0ES analysis for the coupling parameter $\zeta$, the cosmological parameters and the nuisance parameter $M$.}
    \label{fig_kinetic_qso_total}
\end{figure*}

\begin{figure*}[t]
    \centering
    \includegraphics[scale=0.44]{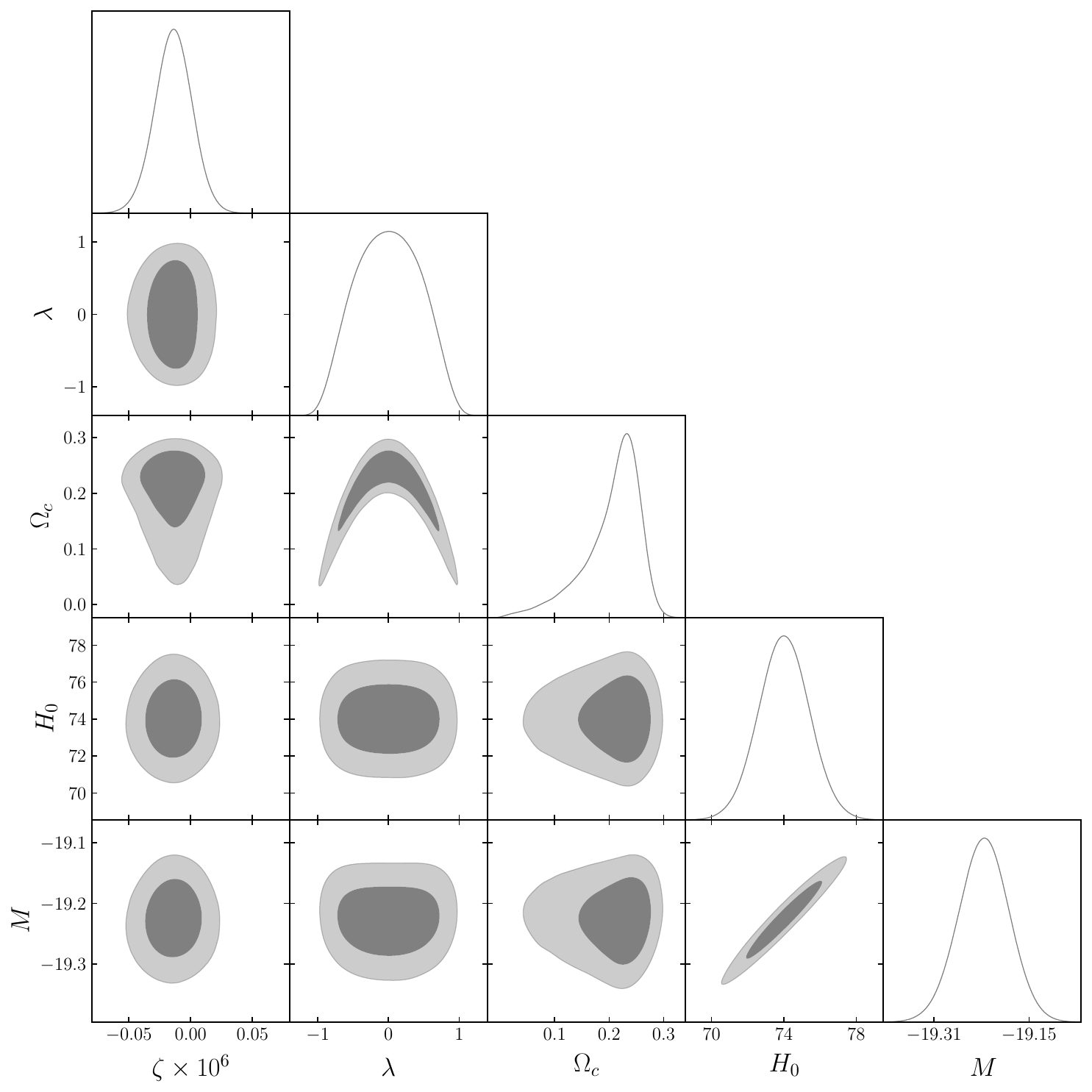}
    \caption{Marginalised contours and posterior distributions (68\% and 95\% confidence levels) of the atomic clocks + Pantheon + SH0ES analysis for the coupling parameter $\zeta$, the cosmological parameters and the nuisance parameter $M$.}
    \label{fig_kinetic_clocks_total}
\end{figure*}

Here we present the plots of the Bayesian analysis for the kinetic model: with Planck (Fig.~\ref{fig_kinetic_planck_ter}), QSO combined with Pantheon and the SH0ES prior (Fig.~\ref{fig_kinetic_qso_total}), and atomic clocks combined with Pantheon and the SH0ES prior (Fig.~\ref{fig_kinetic_clocks_total}).

\clearpage

\bibliography{bib.bib}

\end{document}